\documentclass[manuscript=article]{achemso}
\usepackage{achemso}
\setkeys{acs}{usetitle = true}
\usepackage[version=3]{mhchem} 
\usepackage[T1]{fontenc}       
\usepackage{color,soul}
\usepackage{amsmath,amssymb}
\usepackage{subfig}
\usepackage{natbib}
\usepackage{multirow}
\usepackage[utf8]{inputenc}
\usepackage{amssymb}
\author{Arunima Singh$^{*}$, Preeti Bhumla, Manjari Jain, Saswata Bhattacharya} 
\affiliation{Department of Physics, Indian Institute of Technology Delhi, Hauz Khas, New Delhi:110016, India}
\email{saswata@physics.iitd.ac.in [SB], Arunima.Singh@physics.iitd.ac.in[AS]}
\phone{+91-2659 1359}
\fax{+91-2658 2037}

\title[An \textsf{achemso} demo]
{Electrocatalytic Study for Hydrogen Evolution Reaction on MoS{$_2$}/BP and MoSSe/BP in Acidic Media}


\begin{document}
\begin{abstract} 
\noindent 
Molecular hydrogen (H{$_2$}) production by electrochemical hydrogen evolution reaction (HER) is being actively explored for non-precious-metal based electrocatalysts that are earth-abundant and low cost like MoS{$_2$}. Although it is acid-stable, its applicability is limited by catalytically inactive basal plane, poor electrical transport and inefficient charge transfer at the interface. Therefore, the present work examines its bilayer van der Waals heterostructure (vdW HTS). The second constituent monolayer Boron Phosphide (BP) is advantageous as an electrode material owing to its chemical stability in both oxygen and water environments. Here, we have performed first-principles based calculations under the framework of density functional theory (DFT) for HER in an electrochemical double layer model with the BP monolayer, MoS{$_2$}/BP and  MoSSe/BP vdW HTSs. The climbing image nudged elastic band method (CI-NEB) has been employed to determine the minimum energy pathways for Tafel and Heyrovsky reactions. The calculations yield that Tafel reaction shows no reaction barrier. Thereafter, for Heyrovsky reaction, we have obtained low reaction barrier in the vdW HTSs as compared to that in the BP monolayer. Subsequently, we have observed no significant difference in the reaction profile of MoS{$_2$}/BP and  MoSSe/BP vdW HTSs in case of high coverage (25\%) and 1/3 H{$^+$} concentration (conc.). However, in the case of small coverage (11\%) and 1/3 H{$^+$} conc., MoSSe/BP shows feasible Heyrovsky reaction with no reaction barrier. Finally, on comparing the coverages with 1/4 H{$^+$} conc., we deduce high coverage with low conc. and low coverage with high conc. to be apt for HER via Heyrovsky reaction path. 
\end{abstract}
\section{Introduction}
The availability of clean and renewable energy source governs the tenable development. Innovation in systems like fuel cells, metal-air batteries and water electrolysis positively impacts the environment~\cite{zhang20222d}. The cleanest alternative for the same is the molecular hydrogen (H{$_2$}) and hence, in the present context, we consider materials that support its production~\cite{chu2012opportunities, seh2017combining}. The electrochemical reactions that are in sync with the clean environment aim involve hydrogen oxidation reaction (HOR), oxygen reduction reaction (ORR), hydrogen evolution reaction (HER) and oxygen evolution reaction (OER)~\cite{vij2017nickel}. The former two are associated with fuel cells, while the latter two are associated with water splitting or water electrolysis. There exists wide range of materials that can catalyze these electrochemical reactions by photocatalytic or electrocatalytic pathways~\cite{lu20162d,cao2020metal,jin2021metastable}. The present paper focuses on HER by the electrocatalysts. HER requires large overpotential to be initiated, and therefore catalysts are required to lower the overpotential~\cite{zhang2021transition}.
In this respect, Pt has established itself to be an efficient catalyst~\cite{hansen2021there}. However, its high cost and low abundance have urged the scientific community to find new materials for catalytic applications~\cite{zhao2021non}. In fact, any heterogeneous catalysis under periodic boundary conditions faces the challenge of possessing an apt catalytic material that decreases the reaction barrier~\cite{shi2021electronic}.\\
\noindent HER can occur in both acidic and alkaline media. In either of the media, the reaction steps follow (i) adsorption of H, (ii) its reduction and (iii) desorption as H{$_2$}~\cite{jin2018emerging}. Now, HER has been reported to have sluggish kinetics in alkaline media with ambiguous active sites~\cite{wei2018heterostructured}. Since the electrolytic reactions at the electrode are acidic, we are focusing on the acidic media in the present study. 
The adsorption step is very fast and is termed as the Volmer step~\cite{skulason2007density}: \\
Volmer reaction (fast): H{$^+$} + e{$^-$} $\rightarrow$ H{$_{\textrm{ad}}$}  \\
The subsequent steps take place either as Tafel or Heyrovsky paths. \\
Tafel reaction: 2H{$_{\textrm{ad}}$} $\rightarrow$ H{$_{\textrm{2}}$} \\
Heyrovsky reaction: H{$_{\textrm{ad}}$} + H{$^+$} + e{$^-$} $\rightarrow$ H{$_{\textrm{2}}$} 
\begin{figure}[htb]
	\centering
	\includegraphics[width=0.7\columnwidth,clip]{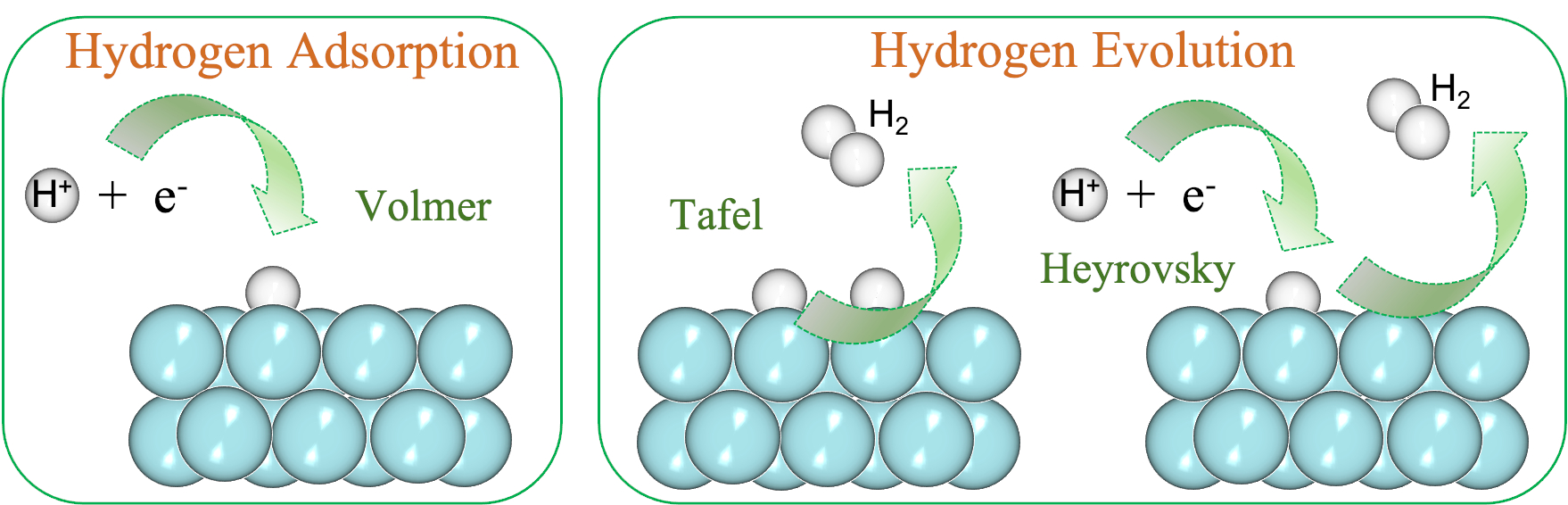}
	\caption{(Color online) HER Steps: Volmer is the adsorption step, and Tafel/Heyrovsky is the evolution step.}
	\label{fig:HER}
\end{figure}
\\
\noindent As previously mentioned, the concept is to obtain material for the reaction that does not include  precious metals like Pt. The literature has shown the transition metals (Fe, Ni, Co), carbides, metal oxides (RuO{$_2$}, IrO{$_2$}), graphene, non-layered 2D materials, metal–organic frameworks (MOFs) and transition metal dichalcogenides (TMDs) as effective HER catalysts~\cite{wang2020two,jin2018emerging, wang2020recent,zhang2021transition,du2021metal, esposito2012new,trasatti1991physical}. We restrict our study to the 2D materials that showcase quantum confinement effects with increased carrier mobility and large surface area~\cite{mir2020recent,C8MH01001C}. This results in their increased catalytically active sites. The monolayer TMDs (in place of graphene) have established themselves as a potent material with optimal band gap suitable for optoelectronics, photocatalysis and electrocatalysis~\cite{ASjpcc,chen2018ACSNano,rao2015comparative,singh2019ACSApplInter,zeng2013optical,wang2015JPCC}. In addition, due to their flexibility, these are widely studied for flexible electronic devices. Literature has reported their use as catalysts for HER, especially on the surface of 1T$'$-MoS{$_2$} and edge sites of 2H-MoS{$_2$}~\cite{wang2019structural,chen2020tuning}. MoS{$_2$} being acid-stable is an added advantage~\cite{tang2016mechanism}. Furthermore, its heterojunctions have also shown promising HER catalytic behaviour~\cite{ASnanoscale,ling2019CompMat,keivanimehr2021electrocatalytic,wei2018heterostructured}. It is pertinent to mention here that the tunability of 2D materials for specific applications is prevalent by defect engineering, strain engineering, stacking order, external field implementation, alloying and forming heterojunctions~\cite{wan2016tuning,zhao2014free,bao2011stacking,feng2012strain,wang2008gate}. Amongst them, formation of heterojunctions with van der Waals forces in between the constituent monolayers are classified under van der Waals heterostructures (vdW HTSs). These have proved a real boon to the field of work because the constituent monolayers retain their properties simultaneously with their combined vdW HTS properties. In addition, the electronegativity difference between the constituent monolayers actuates electron transfer, thereby affecting the HER~\cite{zhao2018heterostructures}. Even if the constituent monolayers have inactive sites, the resulting vdW HTS can be obtained as an active electrocatalyst due to an inbuilt electric field at the interface~\cite{bawari2018hydrogen}.

\noindent Presently, we explore Boron Phosphide (BP) monolayer, MoS{$_2$}/BP and MoSSe/BP vdW HTSs for HER.  Recent works have reported vdW HTSs with BP instead of graphene as it has a similar single atomic layered hexagonal structure, however, along with a band gap~\cite{csahin2009monolayer}. BP monolayer has been reported with low carrier effective mass, high carrier mobility, good mechanical strength, and stability in water environments~\cite{wu2021single,vu2021structural}. Since the lattice parameter of MoS{$_2$} and BP is similar, the MoS{$_2$}/BP vdW HTS becomes a plausible system with minimal lattice mismatch~\cite{mohanta2019interfacing}. BP monolayer has also been synthesized experimentally  ~\cite{padavala2016epitaxy}. In addition, since Janus (MoSSe) has established itself with more catalytically active sites than MoS{$_2$}, we have also analyzed MoSSe/BP vdW HTS. Any prior investigations for HER on these systems are hitherto unknown; hence we have considered these systems for our work.

\noindent The aforementioned HER reaction path should be accounted for the proton and electron free energies. These are incorporated by the computational hydrogen electrode model as proposed by Norskov \textit{et al.}~\cite{skulason2007density}. The model caters to the fundamental problem of large-scale calculation of a real system along with electrolyte by following the electrochemical double layer approach rather than external charge formation. The underlying approximation considers solvated proton upto first bilayer. Until now, no study has been reported with the analysis of vdW HTSs by computational hydrogen electrode model for HER in acidic media to the best of our knowledge~\cite{wiensch2017comparative}. We have initially discussed the stacking configuration and electronic structure.  Subsequently, the computational hydrogen electrode model is discussed. Thereafter, Tafel and Heyrovsky reaction paths are analyzed. Finally, we discuss the electrode potential, and the reaction and activation energies.


\section{Methodology}
\noindent  The first-principles based density functional theory (DFT) calculations have been employed in the present work~\cite{martin2004electronic, martin2016interacting, freysoldt2014RevModPhys, feng2014MaterChemPhys, hohenberg1964PhysRev, kohn1965PhysRev}. The associated code chosen is Vienna \textit{ab initio} simulation package (VASP)~\cite{kresse1996efficient,blochl1994projector,blum2009ComputPhysCommun} with projector augmented wave (PAW) pseudopotentials using plane wave basis. The generalized gradient approximation (GGA) that accounts for the exchange-correlation (xc) interaction amongst electrons is incorporated by PBE xc functional (as proposed by Perdew-Burke-Ernzerhof (PBE)~\cite{stampfl1999PRB, perdew1996PRL}). 
\noindent The Brillouin zone (BZ) sampling of {$2\times2$} K-mesh is used for conjugate gradient minimization with an energy tolerance of 0.001 meV and the force tolerance of 0.001 eV/{\AA}. The intermediate, initial and final energetics are obtained by the BZ sampling of {$6\times6$} K-mesh. The plane wave cut-off energy is set to 500 eV. All the structures are built with 20 {\AA} vacuum that avoids the electrostatic interactions among the periodic images. The two-body Tkatchenko-Scheffler vdW scheme has been employed for obtaining optimized structures~\cite{tkatchenko2009PRL,tkatchenko2012PRL}. This is an iterative scheme based on Hirshfeld partitioning of the electron density. We have employed climbing-image nudged elastic band (CI-NEB) method to obtain minimum energy path for HER~\cite{henkelman2000climbing, henkelman2000improved}. Note that we have not explicitly considered entropic calculations, as in approximation of solvated proton on first layer, 0.2 - 0.3 eV can be added all along the energetics~\cite{skulason2007density}.
In reference to the previous literature, we have not included the spin-orbit coupling (SOC) in our calculations~\cite{Direct_Z_Fu,weng2018honeycomb,ren2020direct}. 


\section{Results and Discussions}
\subsection{Heterostructure }
\noindent 
The present paper features BP monolayer, MoS{$_2$}/BP and MoSSe/BP vdW HTSs for HER assessment.  The lattice parameter of BP monolayer is 3.20 {\AA} and that of MoS{$_2$} is 3.16 \AA. Since the lattice mismatch between them is less (1.2\% as obtained by (l(MoS$_2$) - l(BP))/l(BP), where l(MoS$_2$) and l(BP) is the lattice constant of MoS$_2$ and BP, respectively), the corresponding MoS{$_2$}/BP vdW HTS formed is commensurate~\cite{rahman2018commensurate}. Its corresponding structural and electronic properties are obtained by unit cell configuration (see Fig. S1-S3 in Supplementary Information (SI)), whereby, MoS{$_2$}/BP and MoSSe/BP form type 1 and type 2 alignment and it corroborates with the prior research~\cite{ren2019first,mohanta2019interfacing}. 
\begin{figure}[htb]
	\centering
	\includegraphics[width=0.6\columnwidth,clip]{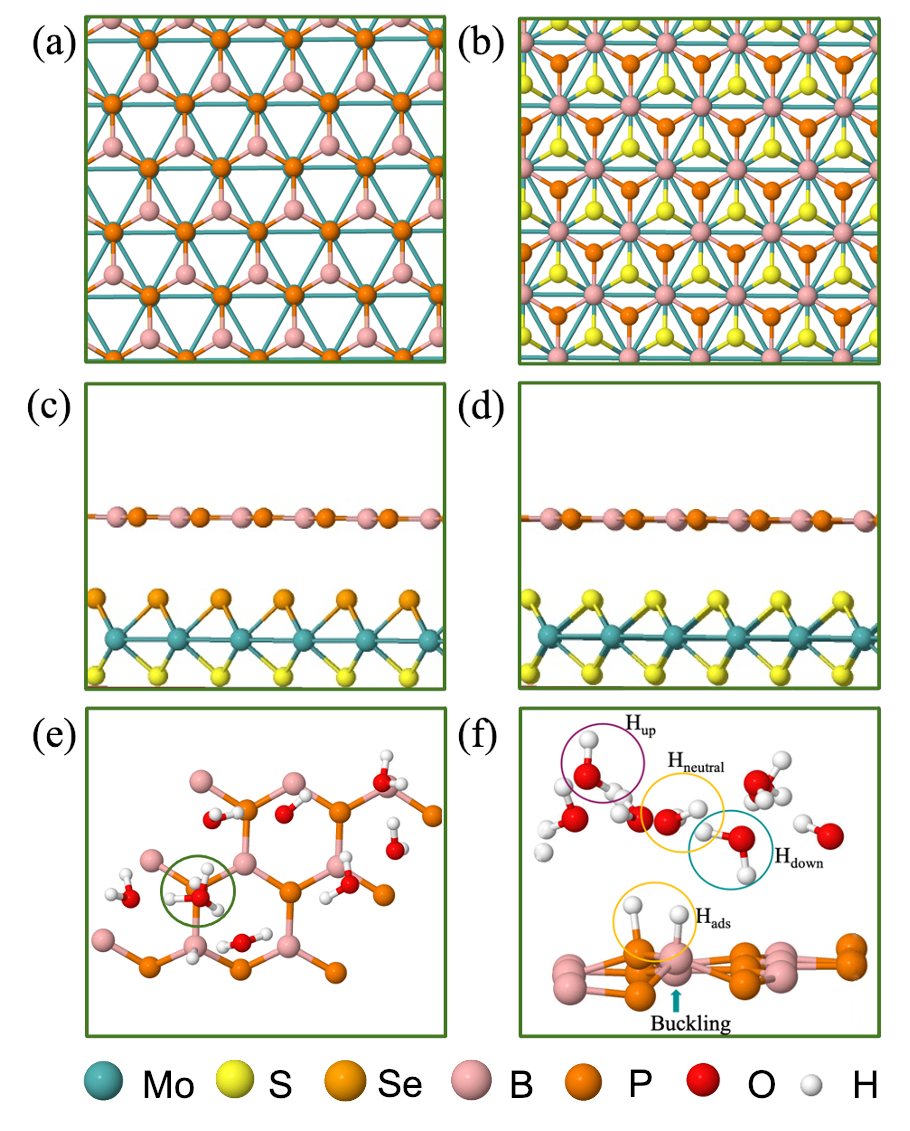}
	\caption{(Color online) (a) and (b) Top view of MoS{$_2$}/BP vdW HTS stacking configurations, (c) and (d) Side view of minimum energy stacking configuration for MoSSe/BP and MoS{$_2$}/BP vdW HTSs, respectively, (e) {$3\times3$} supercell of BP monolayer schematic with 1/3 H{$^+$} conc. i.e., 1 H{$^+$}/3H$_2$O, (f) Water molecule orientations of H$_\textrm{up}$, H$_\textrm{down}$ and H$_\textrm{neutral}$. Buckling on BP can be observed at the H$_\textrm{ads}$ site. }
	\label{fig:Structure}
\end{figure}
Note that, initially, two stacking styles (see Fig.~\ref{fig:Structure} (a) and (b)) between the constituent monolayers were considered, wherein the stacking corresponding to Fig. ~\ref{fig:Structure} (b) has minimum binding energy~\cite{ASnanoscale}. Therefore, we have proceeded with this stacking in our work. 

\subsection{HER Study }
\noindent Now, we advance on HER study, for which we have constituted {$2\times2$} and {$3\times3$} supercells. The former being smaller restricts the proton concentration (conc.) variability, therefore we need larger supercells. In view of this, we are analyzing {$2\times2$} supercell along with {$3\times3$}, because unlike monolayer, the vdW HTS with further large supercell size becomes computationally demanding. The subsequent paragraphs discuss the concepts of coverage and proton conc. for clarity.\\
\noindent The first step is to obtain the coverage that gives $\Delta$G$_\textrm{H}$ $\simeq$ 0 for our study. The number of adsorbed hydrogen (H$_\textrm{ads}$) per surface atoms is defined as the coverage. $\Delta$G$_\textrm{H}$ is the free energy of atomic hydrogen adsorption and is expressed as:
\begin{equation*}
	\Delta \textrm{G}_\textrm{H} = \Delta \textrm{E}_\textrm{H} + \Delta \textrm{E}_\textrm{ZPE} - \textrm{T}\Delta \textrm{S}_\textrm{H} 
\end{equation*}
where,
\begin{equation*}
	\Delta \textrm{E}_\textrm{H} = \textrm{E}[\textrm{nH}] - \textrm{E}[\textrm{(n-1)H}] - 1/2 \textrm{E}[\textrm{H}_\textrm{2}]
\end{equation*}
In the aforementioned equations, $\Delta$E$_\textrm{H}$ is the hydrogen binding energy on the surface of vdW HTS, E[nH] (or E[(n-1)H]) is the energy of the configuration with n (or n-1) number of H$_\textrm{ads}$, $\Delta \textrm{E}_\textrm{ZPE}$ is the zero-point energy of H$_\textrm{ads}$ and $\Delta \textrm{S}_\textrm{H}$ is the  entropy of H$_2$ in the gas phase. At 298 K, $\Delta \textrm{E}_\textrm{ZPE} - \textrm{T}\Delta \textrm{S}_\textrm{H} = \textrm{0.25 eV}$ is well established in literature~\cite{tang2016mechanism}.
We observe {$2\times2$} supercell with 25\% H coverage (2H$_\textrm{ads}$ per 8 surface atoms) and {$3\times3$} supercell with 11\% H coverage (2H$_\textrm{ads}$ per 18 surface atoms) with $\Delta$G$_\textrm{H}$ equal to -0.024 eV and 0.049 eV, respectively. We have deduced these coverages after trials upto 38\%. We have chosen consecutive B and P atomic sites for H$_\textrm{ads}$ as this configuration was found to be the most stable. Also, we observed buckling at the site of H$_\textrm{ads}$ (see Fig. ~\ref{fig:Structure} (f)). 
\begin{figure}[htb]
	\centering
	\includegraphics[width=0.6\columnwidth,clip]{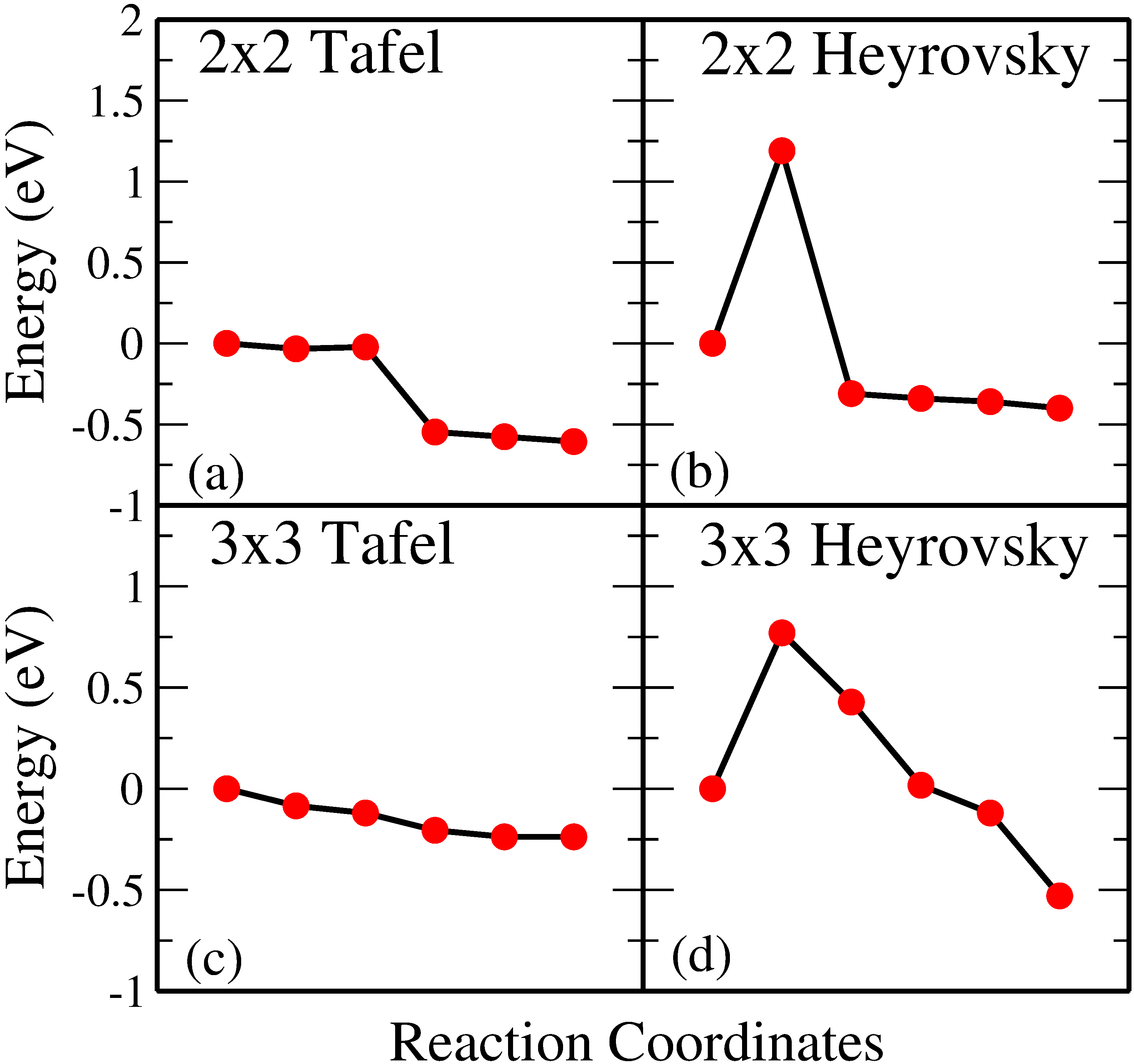}
	\caption{(Color online) {$2\times2$} supercell of BP monolayer showing (a) Tafel reaction path and (b) Heyrovsky reaction path. {$3\times3$} supercell of BP monolayer showing (c) Tafel reaction path and (d) Heyrovsky reaction path.}
	\label{fig:2}
\end{figure}
\\
\noindent We now discuss the optimized systems consisting of water layer (water-solid interface with 3 Å thick water layer) without and with solvated protons (i.e., H{$^+$}). Fig. ~\ref{fig:Structure} (e) shows the BP monolayer (2H$_\textrm{ads}$) with H{$^+$} in {$3\times3$} supercell. Note that the H{$^+$} is in the form of hydronium (H$_3$O) in the water layer. {$2\times2$} supercell is a small supercell and therefore, only 1 H{$^+$} is been considered. However, the corresponding H{$_2$}O molecules in the water layer are varied, thereby constituting 1/3 (i.e., 1 H{$^+$}/3H$_2$O) and 1/4 (i.e., 1 H{$^+$}/4H$_2$O) H{$^+$} conc. 
The configuration corresponding to {$3\times3$} supercell size has been studied for 1/8 (i.e., 1 H{$^+$}/8H$_2$O) H{$^+$} conc. 
\noindent The water orientation (see Fig. ~\ref{fig:Structure} (f)) over the H$_\textrm{ads}$ species is flat and H$_\textrm{up}$ orientation is usually seen on the topmost layer. Further, all H{$_2$}O molecules are not H$_\textrm{down}$, rather, they are at some angular orientations other than strict H$_\textrm{up}$ and H$_\textrm{down}$ configurations. These orientations are essential because the electrostatic potential, as seen from the solid surface, also depends on the same. 
The stability of the vdW HTS along with water layer orientation is established by the similar profile of radial distribution plot at 0K and 300K (see Fig. S6 in SI). 
\subsection{Tafel Reaction Step}
\noindent Fig. ~\ref{fig:2}(a) and ~\ref{fig:2}(b) give Tafel and Heyrovsky reaction steps, respectively on the BP monolayer. This corresponds to the {$2\times2$} supercell with 3 H{$_2$}O molecules and 1/3 H{$^+$} conc., respectively.  
\begin{figure*}[htb]
	\centering
	\includegraphics[width=0.9\textwidth,clip]{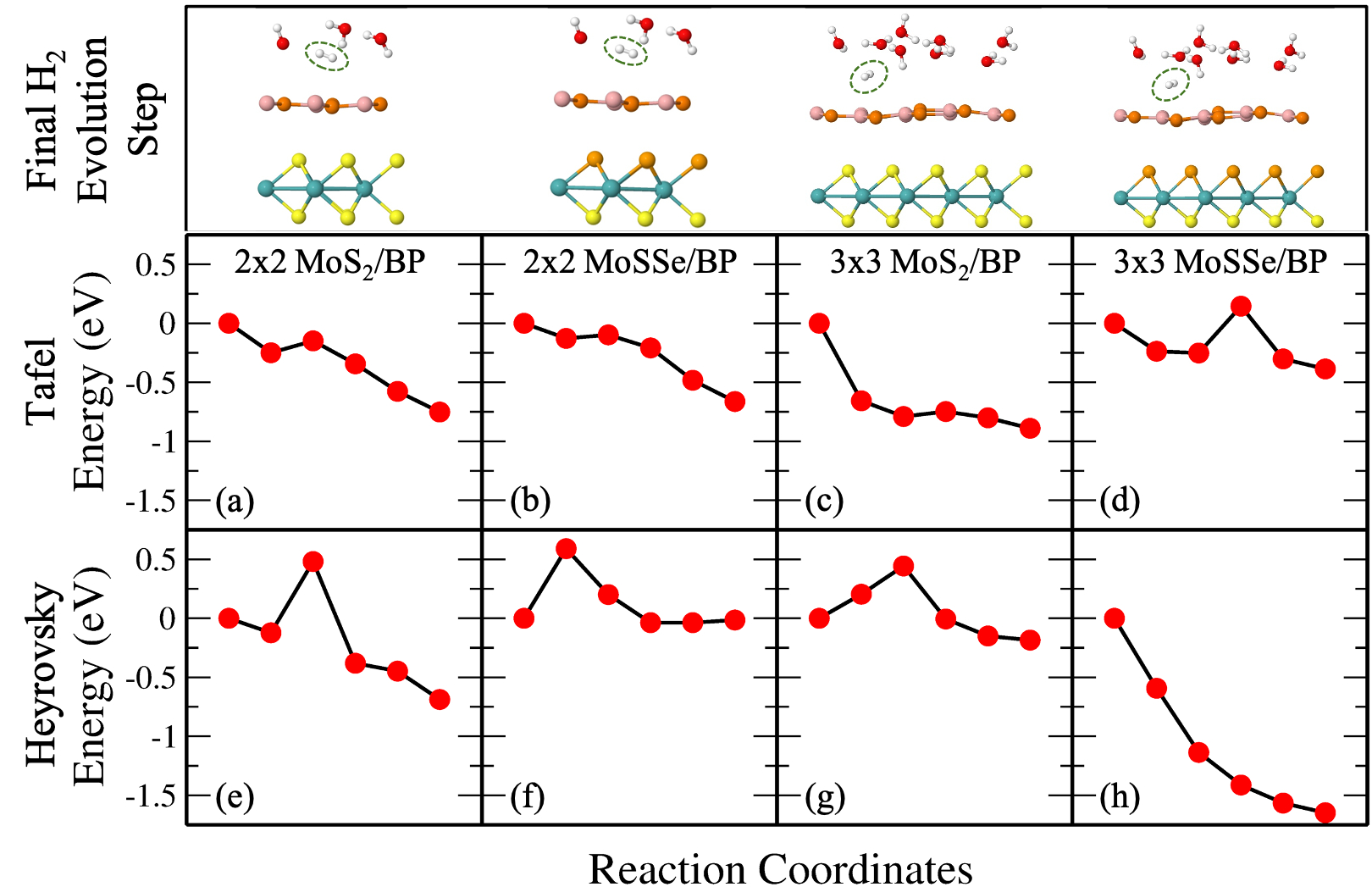}
	\caption{(Color online) (a)-(d) Tafel reaction path (upper row) on MoS{$_2$}/BP  and MoSSe/BP vdW HTSs for $2\times2$ supercell and $3\times3$ supercell. (e)-(f) Heyrovsky reaction path (lower row) on MoS{$_2$}/BP  and MoSSe/BP vdW HTSs for $2\times2$ supercell and $3\times3$ supercell.}
	\label{fig:3}
\end{figure*}
\begin{figure*}[htb]
	\centering
	\includegraphics[width=0.9\textwidth,clip]{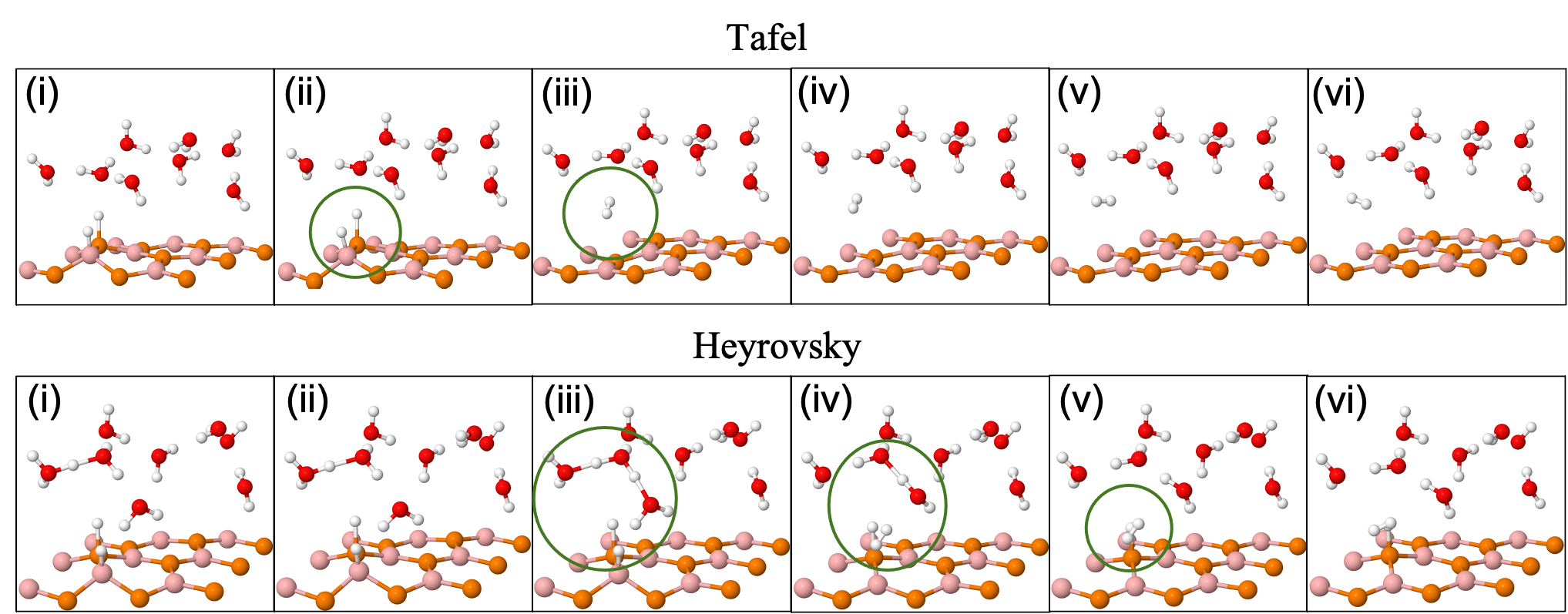}
	\caption{(Color online) Tafel (upper row) and Heyrovsky (lower row) reaction profile snapshots on $3\times3$ BP surface.}
	\label{fig:steps}
\end{figure*}
\\
\noindent The BP monolayer acts as a reference to analyze the reactions for the MoS{$_2$}/BP and MoSSe/BP vdW HTSs. Here, we observe a reaction barrier in Heyrovsky reaction step (1.19 eV) and not in case of the Tafel reaction step. The reaction steps for {$3\times3$} supercell and 1/8 H{$^+$} conc. are given in Fig. ~\ref{fig:2}(c) and ~\ref{fig:2}(d), whereby the  Tafel reaction steps show no barrier and Heyrovsky reaction steps show reduced reaction barrier as compared to that in {$2\times2$} supercell. Further, we first discuss the Tafel reaction step analysis for the vdW HTSs.
\noindent Fig.~\ref{fig:3} highlights the Tafel reaction step on the MoS{$_2$}/BP ((a) and (c)) and MoSSe/BP ((b) and (d)) vdW HTSs. Firstly, no significant difference is observed between MoS{$_2$}/BP and MoSSe/BP vdW HTSs for {$2\times2$} supercell. 
\noindent However, we observe reaction barrier only in case of MoSSe/BP $3\times3$ supercell (0.14 eV).
The overview of the Tafel reaction analysis is consistent with Tafel being surface reaction, thereby, less or no observed reaction barrier. We observed that the minimum energy profile in Tafel reaction is not continuously decreasing; instead, a slight hump is present. This corresponds to the buckling in the BP monolayer. As previously mentioned, the site of H$_\textrm{ads}$ is buckled with respect to other sites, and during the  H$_\textrm{2}$ evolution process, the corresponding BP site adjusts itself to the planar configuration (see Fig.~\ref{fig:steps}). Note that BP surface is considered for the reaction analysis as basal plane of MoS{$_2$} is not catalytically active.
\subsection{Heyrovsky Reaction Step}
\noindent Unlike Tafel, the Heyrovsky reaction step is not a pure surface reaction. It involves charge transfer, thereby affecting the reaction barrier and Fig.~\ref{fig:3} (e)-(h) substantiates the same. We have observed reduction in reaction barrier in the vdW HTSs as compared to that in BP monolayer (refer Fig.~\ref{fig:2} and Fig.~\ref{fig:3}). The $3\times3$ supercell configuration puts forth decreased reaction barrier than in the case of  $2\times2$ supercell configuration. The MoS{$_2$}/BP and MoSSe/BP demonstrates this reduction from 0.48 eV (Fig.~\ref{fig:3} (e)) to 0.43 eV (Fig.~\ref{fig:3} (g)) and 0.59 eV (Fig.~\ref{fig:3} (f)) to 0 eV (Fig.~\ref{fig:3} (h)), respectively. The reduced coverage and hence reduced charge redistribution on the surface can be attributed to the reduced reaction barrier in the $3\times3$ supercell.
Further, we observe significant change in case of MoSSe and this may be attributed to the combined effect of the coverage and the electronegativity difference within the MoSSe layer that affects the charge transfer at the interface.\\
\noindent Apart from the factors that are discussed above, there are structural parameters that affect the reaction steps. The bonds of H in H$_3$O stretch before combining with the H$_\textrm{ads}$. At the transition state, H$_2$ is formed. After that, the atoms adjust themselves to low energy configuration.
After the intermediate step, the B and P atoms adjust, corresponding to H$_\textrm{ads}$, along with the other H{$_2$}O molecules. As in the Tafel scenario, the steps post H{$_2$} formation optimize the H{$_2$} molecule in the water layer. The reaction barrier, therefore, depends on the buckling in the monolayer, the water molecule's orientation, and the coexisting water molecules with H{$^+$} (see Fig.~\ref{fig:steps}).\\
\begin{figure}[htb]
	\centering
	\includegraphics[width=0.75\columnwidth,clip]{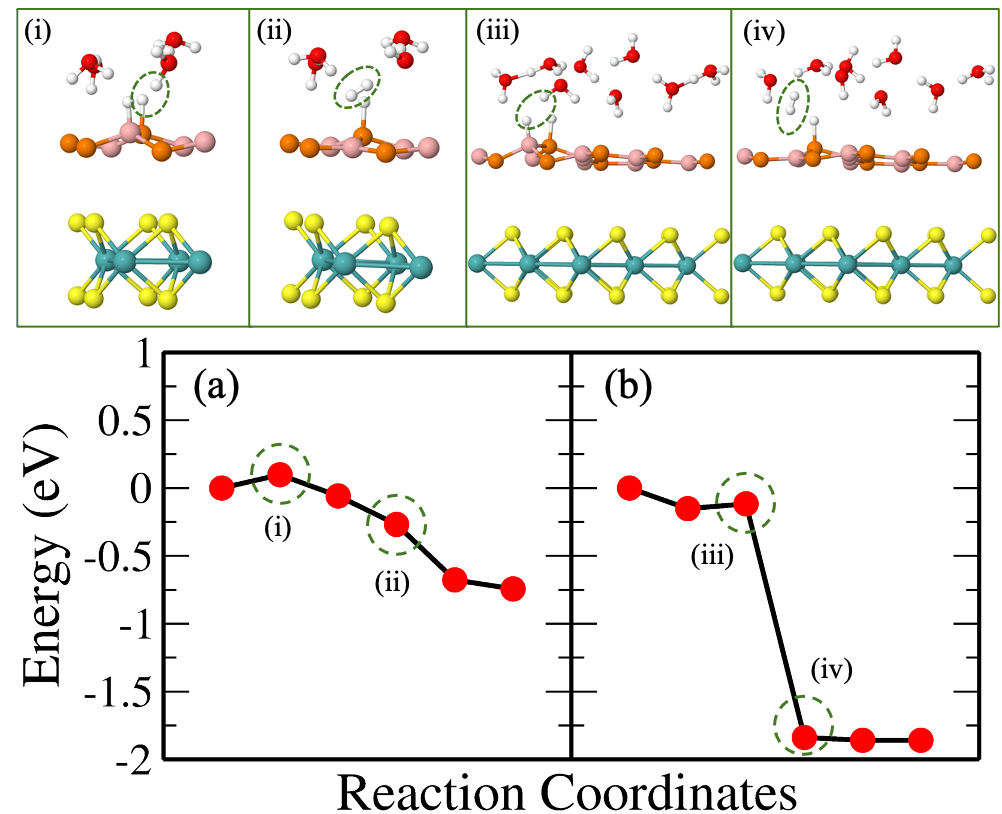}
	\caption{(Color online) Heyrovsky reaction path for MoS{$_2$}/BP vdW HTS with 1/4 H{$^+$} conc. in (a) $2\times2$ supercell and (b) $3\times3$ supercell.}
	\label{fig:6}
\end{figure}
\noindent Finally, we discuss the Heyrovsky reaction in MoS{$_2$}/BP for 1/4 H{$^+$} conc. both in case of $2\times2$ (i.e., 1 H{$^+$}/4H$_2$O) and $3\times3$ (i.e., 2 H{$^+$}/8H$_2$O) supercells. We observed reaction barrier decreases from 0.48 eV (Fig. ~\ref{fig:3} (e)) to 0.09 eV (Fig. ~\ref{fig:6} (a)) and 0.59 eV (Fig. ~\ref{fig:3} (g)) to 0 eV (Fig. ~\ref{fig:6} (b)) in $2\times2$ and $3\times3$ supercells, respectively. This indicates that high coverage prefers low H{$^+$} conc. and vice versa for reduction in reaction barrier. 
\noindent We correlate this with the overpotential of the reaction, as discussed in the following section. Overpotential is the difference between the experimentally obtained reaction potential and the electrode potential. The electrode potential is analyzed only in the Heyrovsky reaction as it involves proton transfer. Therefore, this affects the work function and the potential at which the reaction takes place. 
\subsection{Electrode Potential}
\begin{figure}[htb]
	\centering
	\includegraphics[width=0.8\columnwidth,clip]{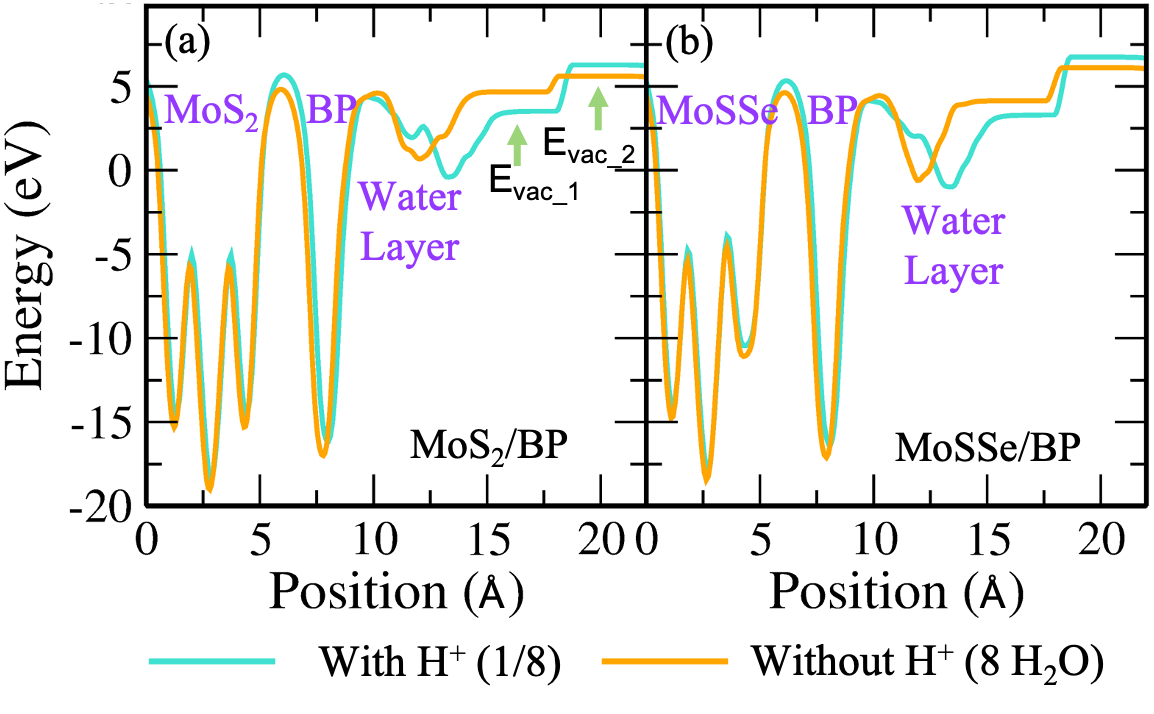}
	\caption{(Color online) Electrostatic potential plot of MoS{$_2$}/BP and MoSSe/BP vdW HTSs depicting water layer with and without H{$^+$}. (a) and (b) $3\times3$ supercell with 1/8 H{$^+$} conc. and 8 H{$_2$}O molecules.}
	\label{fig:overpot}
\end{figure}
\begin{table}[ht]
	\caption{Electrode potential (U) of MoS{$_2$}/BP and MoSSe/BP with and without H{$^+$} in water layer.} 
	
	\begin{tabular}[c]{p{0.20\textwidth}p{0.08\textwidth}p{0.08\textwidth}p{0.08\textwidth}p{0.08\textwidth}}\hline
		\\[-1em]
		vdW HTSs&\multicolumn{2}{c} {With H{$^+$}}&\multicolumn{2}{c}{Without H{$^+$}}\\
		&U$_1$ (V)&U$_2$ (V)&U$_1$ (V)&U$_2$ (V)\\
		\hline
		\\[-1em]
		MoS{$_2$}/BP (2$\times$2)  &   -2.31  &   0.48 & 1.04 & 0.70\\ 
		\\[-1em]
		MoSSe/BP (2$\times$2)   &   -1.83  &   1.07  & -0.68 & 1.31\\ 
		\\[-1em]
		MoS{$_2$}/BP (3$\times$3) &   -2.09 &   -0.05  &-0.09& 0.84\\ 
		\\[-1em]
		MoSSe/BP (3$\times$3)  &  -2.55   &   0.90  &-0.79 & 1.19\\  \hline
	\end{tabular}
	\label{Table1}
	
\end{table}

\noindent The electrode potential (U) of the slab is reported relative to the normal hydrogen electrode (NHE):
\begin{equation*}
	\textrm{U} = \phi - \phi_\textrm{NHE}
\end{equation*} 
Here $\phi$ (E$_\textrm{vac}$ - E$_\textrm{fermi}$) is the work function, and $\phi_\textrm{NHE}$ is taken to be 4.44 eV\cite{skulason2007density,he2018electrochemical,tang2016mechanism}. The work function depends on the surface H coverage, the thickness or number of water bilayers, the water molecule orientation, and the system size. 
In small systems (here $2\times2$), the range of electrode potential analysis is limited to a few H{$^+$} conc. considerations. Fig. ~\ref{fig:overpot} presents the electrostatic potential plot where we have deduced the work function of $3\times3$ MoS{$_2$}/BP and MoSSe/BP. The same for $2\times2$ MoS{$_2$}/BP and MoSSe/BP is shown in Fig. S9 of SI. The potential drops are evident in Fig. ~\ref{fig:overpot}, with a significant drop at the interface of BP and the water layer. The values of U corresponding to water layer with and without H{$^+$} are reported in Table ~\ref{Table1}, which are in the range of -2.5 V to 1.3 V. We have incorporated dipole corrections as the vdW HTSs with two different surfaces maintain two potentials. Moreover, the H$_\textrm{ads}$ and, therefore, the coverage affects the dipole-dipole interactions.  As a result, we report the two values of U, i.e., U$_1$ and U$_2$, corresponding to two vacuum levels of E$_\textrm{vac\_1}$ and E$_\textrm{vac\_2}$, respectively. \\
\noindent As previously discussed the dependence of $\phi$ on water orientation, we have explicitly optimized the H$_\textrm{down}$ configuration for H{$_2$}O molecules. The Heyrovsky reaction path for the same in MoS{$_2$}/BP and MoSSe/BP $3\times3$ supercells can be seen in Fig. S10 (a) and (b) of SI. The obtained barrier is reduced as compared to the $2\times2$ supercells of MoS{$_2$}/BP, MoSSe/BP and $3\times3$ supercell of MoS{$_2$}/BP. The corresponding electrode potential is also reported in Fig. S10 (c) and (d).
\begin{figure}[h]
	\centering
	\includegraphics[width=0.8\columnwidth,clip]{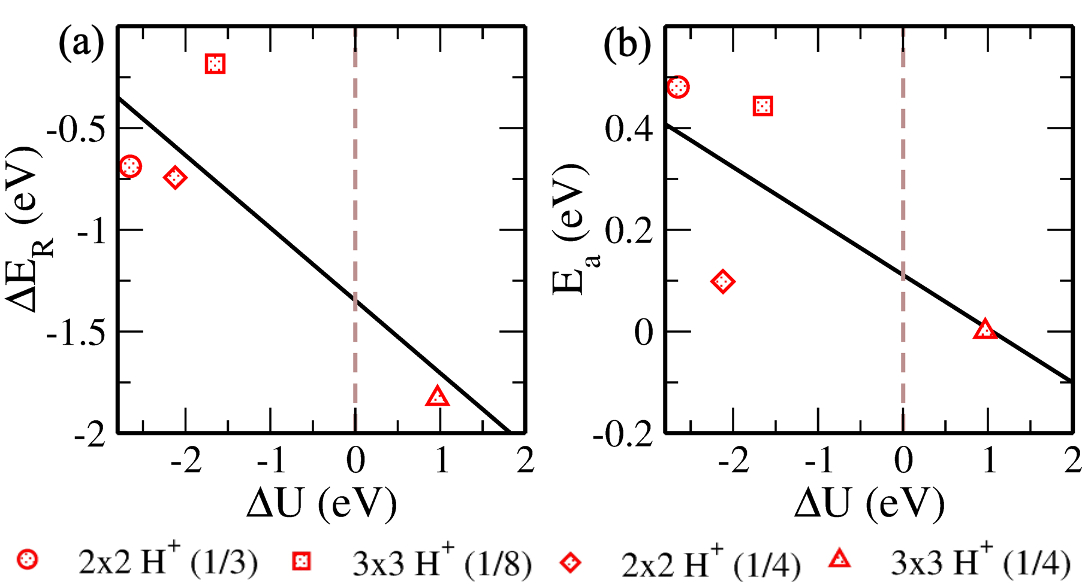}
	\caption{(Color online) Variation of (a) reaction energy ($\Delta \textrm{E}_\textrm{R} = \textrm{E}_\textrm{final}-\textrm{E}_\textrm{initial}$) and (b) activation energy ($\textrm{E}_\textrm{a}$), of configurations with respect to the change in electrode potential ($\Delta \textrm{U} = \textrm{U}_{1_\textrm{initial}}-\textrm{U}_{1_\textrm{final}}$) from initial to final.}
	\label{fig:Rxn}
\end{figure}
\\
\noindent Now we progress towards the extrapolation approach to cater the problem of potential change from initial to final in case of proton transfer Heyrovsky reaction. In this approach, we obtain reaction energies and activation energies of system with different supercell sizes and H{$^+$} conc. The former represents the energy difference between initial and final states, while the latter is the amount of energy required to overcome the reaction barrier. Thereafter we obtain $\Delta \textrm{E}_\textrm{R}$ and $\textrm{E}_\textrm{a}$  vs $\Delta$U plot. The $\Delta$U signifies change in electrode potential from initial to final. Moreover, the change in U$_1$ (corresponding to water layer potential) is significant as compared to the change in U$_2$ (corresponding to MoS{$_2$} layer potential). The potential drop and charge transfer would accordingly affect the U$_1$ and U$_2$. Hence, the reaction taking place at BP layer surface is crucial and we must consider U$_1$ for our analysis of electrode potential. Therefore, the $\Delta$U represented in the plot is corresponding to the U$_1$. On extrapolating the $\Delta \textrm{E}_\textrm{R}$ to $\Delta$U = 0, we obtain -1.33 eV. The negative value indicates the spontaneity of the Heyrovsky reaction step. In case of positive $\Delta \textrm{E}_\textrm{R}$, the Heyrovsky reaction would have been the rate determining step. The corresponding $\textrm{E}_\textrm{a}$ is obtained at 0.11 eV. Hence, on comparing vdW HTS with monolayer, the synergistic effect of the two layers play role in affecting the overpotential and hence the reaction mechanism. 


\section{Conclusion}
\noindent 
In summary, we have modelled a dynamically stable MoS{$_2$}/BP and MoSSe/BP vdW HTSs. They have been studied for HER by computational hydrogen electrode model. The optimized structure with the water layer showed a significant potential drop at the surface-water interface. The electrostatic potential is further affected by the proton solvated in water layer and the H$_\textrm{ads}$ constituting coverage over the surface. {$2\times2$} supercell with 25\% H coverage and {$3\times3$} supercell with 11\% H coverage have been deduced for the calculations. Firstly, the MoS{$_2$}/BP and MoSSe/BP vdW HTSs show reduced barrier height for both Tafel and Heyrovsky reactions in comparison to the BP monolayer. Tafel reaction, being a surface reaction does not require charge transfer, herein corroborates with no or less barrier observed in the MoS{$_2$}/BP and MoSSe/BP vdW HTSs. In case of Heyrovsky reaction, reduced reaction barrier has been reported. The reduced H{$^+$} conc. in small supercell and 25\% coverage substantiates the reduced barrier. Further, there is no significant difference between the MoSSe/BP and MoS{$_2$}/BP vdW HTS, as observed from the minimum energy reaction paths, except in the case of 11\% coverage of MoSSe/BP with no reaction barrier. Hence, the MoSSe based vdW HTS has shown Heyrovsky reaction favoured HER for low coverage. On comparing the supercells (and hence different coverages) with respect to the same H{$^+$}  conc., we observe high coverage to favour low H{$^+$} conc. and vice versa for reduced reaction barrier. Finally, as per the extrapolation approach for $\Delta \textrm{E}_\textrm{R}$ vs $\Delta \textrm{U}$, the Heyrovsky reaction mechanism is established as favourable.


\section{Supporting Information}
 (i) Structural, electronic and optical properties of MoS{$_2$}/BP and MoSSe/BP.
 (ii) Planar averaged charge density plot of MoS{$_2$}/BP and MoSSe/BP.
 (iii) Planar averaged charge density plot of MoS{$_2$}/BP with 1 solvated H{$^+$}.
 (iv) Radial distribution function at 0K and 300K temperatures for MoS{$_2$}/BP configuration.
 (v) $3\times3$ MoS{$_2$}/BP configurations.
 (vi) $3\times3$ MoS{$_2$}/BP configurations with H$_\textrm{down}$ water layer orientation.
 (vii) Electrode potential of $2\times2$ MoS{$_2$}/BP and MoSSe/BP.
 (viii) H$_\textrm{down}$ water orientation reaction path and electrode potential.
 (ix) Tafel reaction steps for $2\times2$ MoS{$_2$}/BP in case of 4H{$_2$}O.
 
 
\section{Acknowledgement}
\noindent \noindent AS acknowledges IIT Delhi for the senior research fellowship. PB acknowledges UGC, India, for the senior research fellowship [1392/(CSIR- UGC NET JUNE 2018)]. MJ acknowledges CSIR, India, for the senior research fellowship [grant no. 09/ 086(1344)/2018-EMR-I]. SB acknowledges the financial support SERB under his Core Research Grant [CRG/2019/000647]. We acknowledge the High Performance Computing (HPC) and Veena Cluster facility at IIT Delhi for computational resources.

\bibliography{ref}

\providecommand{\latin}[1]{#1}
\makeatletter
\providecommand{\doi}
  {\begingroup\let\do\@makeother\dospecials
  \catcode`\{=1 \catcode`\}=2 \doi@aux}
\providecommand{\doi@aux}[1]{\endgroup\texttt{#1}}
\makeatother
\providecommand*\mcitethebibliography{\thebibliography}
\csname @ifundefined\endcsname{endmcitethebibliography}
  {\let\endmcitethebibliography\endthebibliography}{}
\begin{mcitethebibliography}{68}
\providecommand*\natexlab[1]{#1}
\providecommand*\mciteSetBstSublistMode[1]{}
\providecommand*\mciteSetBstMaxWidthForm[2]{}
\providecommand*\mciteBstWouldAddEndPuncttrue
  {\def\EndOfBibitem{\unskip.}}
\providecommand*\mciteBstWouldAddEndPunctfalse
  {\let\EndOfBibitem\relax}
\providecommand*\mciteSetBstMidEndSepPunct[3]{}
\providecommand*\mciteSetBstSublistLabelBeginEnd[3]{}
\providecommand*\EndOfBibitem{}
\mciteSetBstSublistMode{f}
\mciteSetBstMaxWidthForm{subitem}{(\alph{mcitesubitemcount})}
\mciteSetBstSublistLabelBeginEnd
  {\mcitemaxwidthsubitemform\space}
  {\relax}
  {\relax}

\bibitem[Zhang \latin{et~al.}(2022)Zhang, Chen, Chen, and Zhou]{zhang20222d}
Zhang,~X.; Chen,~A.; Chen,~L.; Zhou,~Z. 2D materials bridging experiments and
  computations for electro/photocatalysis. \emph{Advanced Energy Materials}
  \textbf{2022}, \emph{12}, 2003841\relax
\mciteBstWouldAddEndPuncttrue
\mciteSetBstMidEndSepPunct{\mcitedefaultmidpunct}
{\mcitedefaultendpunct}{\mcitedefaultseppunct}\relax
\EndOfBibitem
\bibitem[Chu and Majumdar(2012)Chu, and Majumdar]{chu2012opportunities}
Chu,~S.; Majumdar,~A. Opportunities and challenges for a sustainable energy
  future. \emph{Nature} \textbf{2012}, \emph{488}, 294--303\relax
\mciteBstWouldAddEndPuncttrue
\mciteSetBstMidEndSepPunct{\mcitedefaultmidpunct}
{\mcitedefaultendpunct}{\mcitedefaultseppunct}\relax
\EndOfBibitem
\bibitem[Seh \latin{et~al.}(2017)Seh, Kibsgaard, Dickens, Chorkendorff,
  N{\o}rskov, and Jaramillo]{seh2017combining}
Seh,~Z.~W.; Kibsgaard,~J.; Dickens,~C.~F.; Chorkendorff,~I.; N{\o}rskov,~J.~K.;
  Jaramillo,~T.~F. Combining theory and experiment in electrocatalysis:
  Insights into materials design. \emph{Science} \textbf{2017}, \emph{355},
  4998\relax
\mciteBstWouldAddEndPuncttrue
\mciteSetBstMidEndSepPunct{\mcitedefaultmidpunct}
{\mcitedefaultendpunct}{\mcitedefaultseppunct}\relax
\EndOfBibitem
\bibitem[Vij \latin{et~al.}(2017)Vij, Sultan, Harzandi, Meena, Tiwari, Lee,
  Yoon, and Kim]{vij2017nickel}
Vij,~V.; Sultan,~S.; Harzandi,~A.~M.; Meena,~A.; Tiwari,~J.~N.; Lee,~W.-G.;
  Yoon,~T.; Kim,~K.~S. Nickel-based electrocatalysts for energy-related
  applications: oxygen reduction, oxygen evolution, and hydrogen evolution
  reactions. \emph{ACS Catalysis} \textbf{2017}, \emph{7}, 7196--7225\relax
\mciteBstWouldAddEndPuncttrue
\mciteSetBstMidEndSepPunct{\mcitedefaultmidpunct}
{\mcitedefaultendpunct}{\mcitedefaultseppunct}\relax
\EndOfBibitem
\bibitem[Lu \latin{et~al.}(2016)Lu, Yu, Ma, Chen, and Zhang]{lu20162d}
Lu,~Q.; Yu,~Y.; Ma,~Q.; Chen,~B.; Zhang,~H. 2D
  transition-metal-dichalcogenide-nanosheet-based composites for photocatalytic
  and electrocatalytic hydrogen evolution reactions. \emph{Advanced Materials}
  \textbf{2016}, \emph{28}, 1917--1933\relax
\mciteBstWouldAddEndPuncttrue
\mciteSetBstMidEndSepPunct{\mcitedefaultmidpunct}
{\mcitedefaultendpunct}{\mcitedefaultseppunct}\relax
\EndOfBibitem
\bibitem[Cao and Wang(2020)Cao, and Wang]{cao2020metal}
Cao,~L.; Wang,~C. Metal--Organic Layers for Electrocatalysis and
  Photocatalysis. \emph{ACS Central Science} \textbf{2020}, \emph{6},
  2149--2158\relax
\mciteBstWouldAddEndPuncttrue
\mciteSetBstMidEndSepPunct{\mcitedefaultmidpunct}
{\mcitedefaultendpunct}{\mcitedefaultseppunct}\relax
\EndOfBibitem
\bibitem[Jin \latin{et~al.}(2021)Jin, Song, Paik, and Qiao]{jin2021metastable}
Jin,~H.; Song,~T.; Paik,~U.; Qiao,~S.-Z. Metastable two-dimensional materials
  for electrocatalytic energy conversions. \emph{Accounts of Materials
  Research} \textbf{2021}, \emph{2}, 559--573\relax
\mciteBstWouldAddEndPuncttrue
\mciteSetBstMidEndSepPunct{\mcitedefaultmidpunct}
{\mcitedefaultendpunct}{\mcitedefaultseppunct}\relax
\EndOfBibitem
\bibitem[Zhang \latin{et~al.}(2021)Zhang, Yang, Zhang, Ma, Huang, Li, and
  Wang]{zhang2021transition}
Zhang,~H.; Yang,~X.; Zhang,~H.; Ma,~J.; Huang,~Z.; Li,~J.; Wang,~Y.
  Transition-Metal Carbides as Hydrogen Evolution Reduction Electrocatalysts:
  Synthetic Methods and Optimization Strategies. \emph{Chemistry--A European
  Journal} \textbf{2021}, \emph{27}, 5074--5090\relax
\mciteBstWouldAddEndPuncttrue
\mciteSetBstMidEndSepPunct{\mcitedefaultmidpunct}
{\mcitedefaultendpunct}{\mcitedefaultseppunct}\relax
\EndOfBibitem
\bibitem[Hansen \latin{et~al.}(2021)Hansen, Prats, Toudahl, M{\o}rch~Secher,
  Chan, Kibsgaard, and Chorkendorff]{hansen2021there}
Hansen,~J.~N.; Prats,~H.; Toudahl,~K.~K.; M{\o}rch~Secher,~N.; Chan,~K.;
  Kibsgaard,~J.; Chorkendorff,~I. Is there anything better than Pt for HER?
  \emph{ACS Energy Letters} \textbf{2021}, \emph{6}, 1175--1180\relax
\mciteBstWouldAddEndPuncttrue
\mciteSetBstMidEndSepPunct{\mcitedefaultmidpunct}
{\mcitedefaultendpunct}{\mcitedefaultseppunct}\relax
\EndOfBibitem
\bibitem[Zhao \latin{et~al.}(2021)Zhao, Chen, Sun, and Pan]{zhao2021non}
Zhao,~G.; Chen,~J.; Sun,~W.; Pan,~H. Non-Platinum Group Metal Electrocatalysts
  toward Efficient Hydrogen Oxidation Reaction. \emph{Advanced Functional
  Materials} \textbf{2021}, \emph{31}, 2010633\relax
\mciteBstWouldAddEndPuncttrue
\mciteSetBstMidEndSepPunct{\mcitedefaultmidpunct}
{\mcitedefaultendpunct}{\mcitedefaultseppunct}\relax
\EndOfBibitem
\bibitem[Shi \latin{et~al.}(2021)Shi, Ma, Xiao, Yin, Huang, Huang, Zheng, Mu,
  Huang, Shi, \latin{et~al.} others]{shi2021electronic}
Shi,~Y.; Ma,~Z.-R.; Xiao,~Y.-Y.; Yin,~Y.-C.; Huang,~W.-M.; Huang,~Z.-C.;
  Zheng,~Y.-Z.; Mu,~F.-Y.; Huang,~R.; Shi,~G.-Y., \latin{et~al.}  Electronic
  metal--support interaction modulates single-atom platinum catalysis for
  hydrogen evolution reaction. \emph{Nature Communications} \textbf{2021},
  \emph{12}, 1--11\relax
\mciteBstWouldAddEndPuncttrue
\mciteSetBstMidEndSepPunct{\mcitedefaultmidpunct}
{\mcitedefaultendpunct}{\mcitedefaultseppunct}\relax
\EndOfBibitem
\bibitem[Jin \latin{et~al.}(2018)Jin, Guo, Liu, Liu, Vasileff, Jiao, Zheng, and
  Qiao]{jin2018emerging}
Jin,~H.; Guo,~C.; Liu,~X.; Liu,~J.; Vasileff,~A.; Jiao,~Y.; Zheng,~Y.;
  Qiao,~S.-Z. Emerging two-dimensional nanomaterials for electrocatalysis.
  \emph{Chemical Reviews} \textbf{2018}, \emph{118}, 6337--6408\relax
\mciteBstWouldAddEndPuncttrue
\mciteSetBstMidEndSepPunct{\mcitedefaultmidpunct}
{\mcitedefaultendpunct}{\mcitedefaultseppunct}\relax
\EndOfBibitem
\bibitem[Wei \latin{et~al.}(2018)Wei, Zhou, Long, Xue, Liao, Wei, and
  Xu]{wei2018heterostructured}
Wei,~J.; Zhou,~M.; Long,~A.; Xue,~Y.; Liao,~H.; Wei,~C.; Xu,~Z.~J.
  Heterostructured electrocatalysts for hydrogen evolution reaction under
  alkaline conditions. \emph{Nano-Micro Letters} \textbf{2018}, \emph{10},
  1--15\relax
\mciteBstWouldAddEndPuncttrue
\mciteSetBstMidEndSepPunct{\mcitedefaultmidpunct}
{\mcitedefaultendpunct}{\mcitedefaultseppunct}\relax
\EndOfBibitem
\bibitem[Sk{\'u}lason \latin{et~al.}(2007)Sk{\'u}lason, Karlberg, Rossmeisl,
  Bligaard, Greeley, J{\'o}nsson, and N{\o}rskov]{skulason2007density}
Sk{\'u}lason,~E.; Karlberg,~G.~S.; Rossmeisl,~J.; Bligaard,~T.; Greeley,~J.;
  J{\'o}nsson,~H.; N{\o}rskov,~J.~K. Density functional theory calculations for
  the hydrogen evolution reaction in an electrochemical double layer on the Pt
  (111) electrode. \emph{Physical Chemistry Chemical Physics} \textbf{2007},
  \emph{9}, 3241--3250\relax
\mciteBstWouldAddEndPuncttrue
\mciteSetBstMidEndSepPunct{\mcitedefaultmidpunct}
{\mcitedefaultendpunct}{\mcitedefaultseppunct}\relax
\EndOfBibitem
\bibitem[Wang \latin{et~al.}(2020)Wang, Zhang, Mao, and Wang]{wang2020two}
Wang,~Y.; Zhang,~Z.; Mao,~Y.; Wang,~X. Two-dimensional nonlayered materials for
  electrocatalysis. \emph{Energy \& Environmental Science} \textbf{2020},
  \emph{13}, 3993--4016\relax
\mciteBstWouldAddEndPuncttrue
\mciteSetBstMidEndSepPunct{\mcitedefaultmidpunct}
{\mcitedefaultendpunct}{\mcitedefaultseppunct}\relax
\EndOfBibitem
\bibitem[Wang \latin{et~al.}(2020)Wang, Zhang, Yang, Ma, Huang, and
  Li]{wang2020recent}
Wang,~Y.; Zhang,~H.; Yang,~X.; Ma,~J.; Huang,~Z.; Li,~J. Recent advances in
  transition-metal carbides: from controlled preparation to hydrogen evolution
  reaction application. \emph{Chemistry (Weinheim an der Bergstrasse, Germany)}
  \textbf{2020}, \emph{118}, 5074--5090\relax
\mciteBstWouldAddEndPuncttrue
\mciteSetBstMidEndSepPunct{\mcitedefaultmidpunct}
{\mcitedefaultendpunct}{\mcitedefaultseppunct}\relax
\EndOfBibitem
\bibitem[Du \latin{et~al.}(2021)Du, Li, and Sun]{du2021metal}
Du,~J.; Li,~F.; Sun,~L. Metal--organic frameworks and their derivatives as
  electrocatalysts for the oxygen evolution reaction. \emph{Chemical Society
  Reviews} \textbf{2021}, \emph{50}, 2663--2695\relax
\mciteBstWouldAddEndPuncttrue
\mciteSetBstMidEndSepPunct{\mcitedefaultmidpunct}
{\mcitedefaultendpunct}{\mcitedefaultseppunct}\relax
\EndOfBibitem
\bibitem[Esposito \latin{et~al.}(2012)Esposito, Hunt, Kimmel, and
  Chen]{esposito2012new}
Esposito,~D.~V.; Hunt,~S.~T.; Kimmel,~Y.~C.; Chen,~J.~G. A new class of
  electrocatalysts for hydrogen production from water electrolysis: metal
  monolayers supported on low-cost transition metal carbides. \emph{Journal of
  the American Chemical Society} \textbf{2012}, \emph{134}, 3025--3033\relax
\mciteBstWouldAddEndPuncttrue
\mciteSetBstMidEndSepPunct{\mcitedefaultmidpunct}
{\mcitedefaultendpunct}{\mcitedefaultseppunct}\relax
\EndOfBibitem
\bibitem[Trasatti(1991)]{trasatti1991physical}
Trasatti,~S. Physical electrochemistry of ceramic oxides. \emph{Electrochimica
  Acta} \textbf{1991}, \emph{36}, 225--241\relax
\mciteBstWouldAddEndPuncttrue
\mciteSetBstMidEndSepPunct{\mcitedefaultmidpunct}
{\mcitedefaultendpunct}{\mcitedefaultseppunct}\relax
\EndOfBibitem
\bibitem[Mir \latin{et~al.}(2020)Mir, Yadav, and Singh]{mir2020recent}
Mir,~S.~H.; Yadav,~V.~K.; Singh,~J.~K. Recent advances in the carrier mobility
  of two-dimensional materials: a theoretical perspective. \emph{ACS Omega}
  \textbf{2020}, \emph{5}, 14203--14211\relax
\mciteBstWouldAddEndPuncttrue
\mciteSetBstMidEndSepPunct{\mcitedefaultmidpunct}
{\mcitedefaultendpunct}{\mcitedefaultseppunct}\relax
\EndOfBibitem
\bibitem[Zhang \latin{et~al.}(2018)Zhang, Wang, Niu, Li, Chen, and
  Wang]{C8MH01001C}
Zhang,~X.; Wang,~B.; Niu,~X.; Li,~Y.; Chen,~Y.; Wang,~J. Bi2OS2: a direct-gap
  two-dimensional semiconductor with high carrier mobility and surface electron
  states. \emph{Mater. Horiz.} \textbf{2018}, \emph{5}, 1058--1064\relax
\mciteBstWouldAddEndPuncttrue
\mciteSetBstMidEndSepPunct{\mcitedefaultmidpunct}
{\mcitedefaultendpunct}{\mcitedefaultseppunct}\relax
\EndOfBibitem
\bibitem[Singh \latin{et~al.}(2020)Singh, Basera, Saini, Kumar, and
  Bhattacharya]{ASjpcc}
Singh,~A.; Basera,~P.; Saini,~S.; Kumar,~M.; Bhattacharya,~S. Importance of
  Many-Body Dispersion in the Stability of Vacancies and Antisites in
  Free-Standing Monolayer of MoS{$_2$} from First-Principles Approaches.
  \emph{The Journal of Physical Chemistry C} \textbf{2020}, \emph{124},
  1390--1397\relax
\mciteBstWouldAddEndPuncttrue
\mciteSetBstMidEndSepPunct{\mcitedefaultmidpunct}
{\mcitedefaultendpunct}{\mcitedefaultseppunct}\relax
\EndOfBibitem
\bibitem[Chen \latin{et~al.}(2018)Chen, Huang, Ji, Adepalli, Yin, Ling, Wang,
  Xue, Dresselhaus, Kong, \latin{et~al.} others]{chen2018ACSNano}
Chen,~Y.; Huang,~S.; Ji,~X.; Adepalli,~K.; Yin,~K.; Ling,~X.; Wang,~X.;
  Xue,~J.; Dresselhaus,~M.; Kong,~J., \latin{et~al.}  Tuning electronic
  structure of single layer MoS{$_2$} through defect and interface engineering.
  \emph{ACS Nano} \textbf{2018}, \emph{12}, 2569--2579\relax
\mciteBstWouldAddEndPuncttrue
\mciteSetBstMidEndSepPunct{\mcitedefaultmidpunct}
{\mcitedefaultendpunct}{\mcitedefaultseppunct}\relax
\EndOfBibitem
\bibitem[Rao \latin{et~al.}(2015)Rao, Gopalakrishnan, and
  Maitra]{rao2015comparative}
Rao,~C.; Gopalakrishnan,~K.; Maitra,~U. Comparative study of potential
  applications of graphene, MoS{$_2$} , and other two-dimensional materials in
  energy devices, sensors, and related areas. \emph{ACS Applied Materials \&
  Interfaces} \textbf{2015}, \emph{7}, 7809--7832\relax
\mciteBstWouldAddEndPuncttrue
\mciteSetBstMidEndSepPunct{\mcitedefaultmidpunct}
{\mcitedefaultendpunct}{\mcitedefaultseppunct}\relax
\EndOfBibitem
\bibitem[Singh \latin{et~al.}(2019)Singh, Singh, Kim, Yeom, and
  Nalwa]{singh2019ACSApplInter}
Singh,~E.; Singh,~P.; Kim,~K.~S.; Yeom,~G.~Y.; Nalwa,~H.~S. Flexible molybdenum
  disulfide (MoS2) atomic layers for wearable electronics and optoelectronics.
  \emph{ACS Applied Materials \& Interfaces} \textbf{2019}, \emph{11},
  11061--11105\relax
\mciteBstWouldAddEndPuncttrue
\mciteSetBstMidEndSepPunct{\mcitedefaultmidpunct}
{\mcitedefaultendpunct}{\mcitedefaultseppunct}\relax
\EndOfBibitem
\bibitem[Zeng \latin{et~al.}(2013)Zeng, Liu, Dai, Yan, Zhu, He, Xie, Xu, Chen,
  Yao, \latin{et~al.} others]{zeng2013optical}
Zeng,~H.; Liu,~G.-B.; Dai,~J.; Yan,~Y.; Zhu,~B.; He,~R.; Xie,~L.; Xu,~S.;
  Chen,~X.; Yao,~W., \latin{et~al.}  Optical signature of symmetry variations
  and spin-valley coupling in atomically thin tungsten dichalcogenides.
  \emph{Scientific Reports} \textbf{2013}, \emph{3}, 1608\relax
\mciteBstWouldAddEndPuncttrue
\mciteSetBstMidEndSepPunct{\mcitedefaultmidpunct}
{\mcitedefaultendpunct}{\mcitedefaultseppunct}\relax
\EndOfBibitem
\bibitem[Wang and Guo(2015)Wang, and Guo]{wang2015JPCC}
Wang,~C.-Y.; Guo,~G.-Y. Nonlinear optical properties of transition-metal
  dichalcogenide MX2 (M= Mo, W; X= S, Se) monolayers and trilayers from
  first-principles calculations. \emph{The Journal of Physical Chemistry C}
  \textbf{2015}, \emph{119}, 13268--13276\relax
\mciteBstWouldAddEndPuncttrue
\mciteSetBstMidEndSepPunct{\mcitedefaultmidpunct}
{\mcitedefaultendpunct}{\mcitedefaultseppunct}\relax
\EndOfBibitem
\bibitem[Wang \latin{et~al.}(2019)Wang, Xiao, Liu, Chiang, Kuai, Peng, Lin,
  Meng, Zhao, Choi, \latin{et~al.} others]{wang2019structural}
Wang,~H.; Xiao,~X.; Liu,~S.; Chiang,~C.-L.; Kuai,~X.; Peng,~C.-K.; Lin,~Y.-C.;
  Meng,~X.; Zhao,~J.; Choi,~J., \latin{et~al.}  Structural and electronic
  optimization of MoS2 edges for hydrogen evolution. \emph{Journal of the
  American Chemical Society} \textbf{2019}, \emph{141}, 18578--18584\relax
\mciteBstWouldAddEndPuncttrue
\mciteSetBstMidEndSepPunct{\mcitedefaultmidpunct}
{\mcitedefaultendpunct}{\mcitedefaultseppunct}\relax
\EndOfBibitem
\bibitem[Chen \latin{et~al.}(2020)Chen, Li, Tang, and Tang]{chen2020tuning}
Chen,~J.; Li,~F.; Tang,~Y.; Tang,~Q. Tuning the phase stability and surface HER
  activity of 1T$^\slash$-MoS$_2$ by covalent chemical functionalization.
  \emph{Journal of Materials Chemistry C} \textbf{2020}, \emph{8},
  15852--15859\relax
\mciteBstWouldAddEndPuncttrue
\mciteSetBstMidEndSepPunct{\mcitedefaultmidpunct}
{\mcitedefaultendpunct}{\mcitedefaultseppunct}\relax
\EndOfBibitem
\bibitem[Tang and Jiang(2016)Tang, and Jiang]{tang2016mechanism}
Tang,~Q.; Jiang,~D.-e. Mechanism of hydrogen evolution reaction on 1T-MoS2 from
  first principles. \emph{ACS Catalysis} \textbf{2016}, \emph{6},
  4953--4961\relax
\mciteBstWouldAddEndPuncttrue
\mciteSetBstMidEndSepPunct{\mcitedefaultmidpunct}
{\mcitedefaultendpunct}{\mcitedefaultseppunct}\relax
\EndOfBibitem
\bibitem[Singh \latin{et~al.}(2021)Singh, Jain, and Bhattacharya]{ASnanoscale}
Singh,~A.; Jain,~M.; Bhattacharya,~S. MoS{$_2$} and Janus (MoSSe) based 2D van
  der Waals heterostructures: emerging direct Z-scheme photocatalysts.
  \emph{Nanoscale Advances} \textbf{2021}, \emph{3}, 2837--2845\relax
\mciteBstWouldAddEndPuncttrue
\mciteSetBstMidEndSepPunct{\mcitedefaultmidpunct}
{\mcitedefaultendpunct}{\mcitedefaultseppunct}\relax
\EndOfBibitem
\bibitem[Ling \latin{et~al.}(2019)Ling, Kang, Jing, Zeng, Chen, Liu, Zhang, Qi,
  Fang, and Zhou]{ling2019CompMat}
Ling,~F.; Kang,~W.; Jing,~H.; Zeng,~W.; Chen,~Y.; Liu,~X.; Zhang,~Y.; Qi,~L.;
  Fang,~L.; Zhou,~M. Enhancing hydrogen evolution on the basal plane of
  transition metal dichacolgenide van der Waals heterostructures. \emph{NPJ
  Computational Materials} \textbf{2019}, \emph{5}, 20\relax
\mciteBstWouldAddEndPuncttrue
\mciteSetBstMidEndSepPunct{\mcitedefaultmidpunct}
{\mcitedefaultendpunct}{\mcitedefaultseppunct}\relax
\EndOfBibitem
\bibitem[Keivanimehr \latin{et~al.}(2021)Keivanimehr, Habibzadeh, Baghban,
  Esmaeili, Mohaddespour, Mashhadzadeh, Ganjali, Saeb, Fierro, and
  Celzard]{keivanimehr2021electrocatalytic}
Keivanimehr,~F.; Habibzadeh,~S.; Baghban,~A.; Esmaeili,~A.; Mohaddespour,~A.;
  Mashhadzadeh,~A.~H.; Ganjali,~M.~R.; Saeb,~M.~R.; Fierro,~V.; Celzard,~A.
  Electrocatalytic hydrogen evolution on the noble metal-free MoS2/carbon
  nanotube heterostructure: a theoretical study. \emph{Scientific Reports}
  \textbf{2021}, \emph{11}, 1--9\relax
\mciteBstWouldAddEndPuncttrue
\mciteSetBstMidEndSepPunct{\mcitedefaultmidpunct}
{\mcitedefaultendpunct}{\mcitedefaultseppunct}\relax
\EndOfBibitem
\bibitem[Wan \latin{et~al.}(2016)Wan, Lacey, Dai, Bao, Fuhrer, and
  Hu]{wan2016tuning}
Wan,~J.; Lacey,~S.~D.; Dai,~J.; Bao,~W.; Fuhrer,~M.~S.; Hu,~L. Tuning
  two-dimensional nanomaterials by intercalation: materials, properties and
  applications. \emph{Chemical Society Reviews} \textbf{2016}, \emph{45},
  6742--6765\relax
\mciteBstWouldAddEndPuncttrue
\mciteSetBstMidEndSepPunct{\mcitedefaultmidpunct}
{\mcitedefaultendpunct}{\mcitedefaultseppunct}\relax
\EndOfBibitem
\bibitem[Zhao \latin{et~al.}(2014)Zhao, Deng, Bachmatiuk, Sandeep, Popov,
  Eckert, and R{\"u}mmeli]{zhao2014free}
Zhao,~J.; Deng,~Q.; Bachmatiuk,~A.; Sandeep,~G.; Popov,~A.; Eckert,~J.;
  R{\"u}mmeli,~M.~H. Free-standing single-atom-thick iron membranes suspended
  in graphene pores. \emph{Science} \textbf{2014}, \emph{343}, 1228--1232\relax
\mciteBstWouldAddEndPuncttrue
\mciteSetBstMidEndSepPunct{\mcitedefaultmidpunct}
{\mcitedefaultendpunct}{\mcitedefaultseppunct}\relax
\EndOfBibitem
\bibitem[Bao \latin{et~al.}(2011)Bao, Jing, Velasco, Lee, Liu, Tran, Standley,
  Aykol, Cronin, Smirnov, \latin{et~al.} others]{bao2011stacking}
Bao,~W.; Jing,~L.; Velasco,~J.; Lee,~Y.; Liu,~G.; Tran,~D.; Standley,~B.;
  Aykol,~M.; Cronin,~S.; Smirnov,~D., \latin{et~al.}  Stacking-dependent band
  gap and quantum transport in trilayer graphene. \emph{Nature Physics}
  \textbf{2011}, \emph{7}, 948--952\relax
\mciteBstWouldAddEndPuncttrue
\mciteSetBstMidEndSepPunct{\mcitedefaultmidpunct}
{\mcitedefaultendpunct}{\mcitedefaultseppunct}\relax
\EndOfBibitem
\bibitem[Feng \latin{et~al.}(2012)Feng, Qian, Huang, and Li]{feng2012strain}
Feng,~J.; Qian,~X.; Huang,~C.-W.; Li,~J. Strain-engineered artificial atom as a
  broad-spectrum solar energy funnel. \emph{Nature Photonics} \textbf{2012},
  \emph{6}, 866--872\relax
\mciteBstWouldAddEndPuncttrue
\mciteSetBstMidEndSepPunct{\mcitedefaultmidpunct}
{\mcitedefaultendpunct}{\mcitedefaultseppunct}\relax
\EndOfBibitem
\bibitem[Wang \latin{et~al.}(2008)Wang, Zhang, Tian, Girit, Zettl, Crommie, and
  Shen]{wang2008gate}
Wang,~F.; Zhang,~Y.; Tian,~C.; Girit,~C.; Zettl,~A.; Crommie,~M.; Shen,~Y.~R.
  Gate-variable optical transitions in graphene. \emph{Science} \textbf{2008},
  \emph{320}, 206--209\relax
\mciteBstWouldAddEndPuncttrue
\mciteSetBstMidEndSepPunct{\mcitedefaultmidpunct}
{\mcitedefaultendpunct}{\mcitedefaultseppunct}\relax
\EndOfBibitem
\bibitem[Zhao \latin{et~al.}(2018)Zhao, Rui, Dou, and
  Sun]{zhao2018heterostructures}
Zhao,~G.; Rui,~K.; Dou,~S.~X.; Sun,~W. Heterostructures for electrochemical
  hydrogen evolution reaction: a review. \emph{Advanced Functional Materials}
  \textbf{2018}, \emph{28}, 1803291\relax
\mciteBstWouldAddEndPuncttrue
\mciteSetBstMidEndSepPunct{\mcitedefaultmidpunct}
{\mcitedefaultendpunct}{\mcitedefaultseppunct}\relax
\EndOfBibitem
\bibitem[Bawari \latin{et~al.}(2018)Bawari, Kaley, Pal, Vineesh, Ghosh, Mondal,
  and Narayanan]{bawari2018hydrogen}
Bawari,~S.; Kaley,~N.~M.; Pal,~S.; Vineesh,~T.~V.; Ghosh,~S.; Mondal,~J.;
  Narayanan,~T.~N. On the hydrogen evolution reaction activity of graphene--hBN
  van der Waals heterostructures. \emph{Physical Chemistry Chemical Physics}
  \textbf{2018}, \emph{20}, 15007--15014\relax
\mciteBstWouldAddEndPuncttrue
\mciteSetBstMidEndSepPunct{\mcitedefaultmidpunct}
{\mcitedefaultendpunct}{\mcitedefaultseppunct}\relax
\EndOfBibitem
\bibitem[{\c{S}}ahin \latin{et~al.}(2009){\c{S}}ahin, Cahangirov, Topsakal,
  Bekaroglu, Akturk, Senger, and Ciraci]{csahin2009monolayer}
{\c{S}}ahin,~H.; Cahangirov,~S.; Topsakal,~M.; Bekaroglu,~E.; Akturk,~E.;
  Senger,~R.~T.; Ciraci,~S. Monolayer honeycomb structures of group-IV elements
  and III-V binary compounds: First-principles calculations. \emph{Physical
  Review B} \textbf{2009}, \emph{80}, 155453\relax
\mciteBstWouldAddEndPuncttrue
\mciteSetBstMidEndSepPunct{\mcitedefaultmidpunct}
{\mcitedefaultendpunct}{\mcitedefaultseppunct}\relax
\EndOfBibitem
\bibitem[Wu \latin{et~al.}(2021)Wu, Li, and Yu]{wu2021single}
Wu,~J.; Li,~J.-H.; Yu,~Y.-X. Single Nb or W atom-embedded BP monolayers as
  highly selective and stable electrocatalysts for nitrogen fixation with
  low-onset potentials. \emph{ACS Applied Materials \& Interfaces}
  \textbf{2021}, \emph{13}, 10026--10036\relax
\mciteBstWouldAddEndPuncttrue
\mciteSetBstMidEndSepPunct{\mcitedefaultmidpunct}
{\mcitedefaultendpunct}{\mcitedefaultseppunct}\relax
\EndOfBibitem
\bibitem[Vu \latin{et~al.}(2021)Vu, Kartamyshev, Hieu, Dang, Nguyen, Poklonski,
  Nguyen, Phuc, and Hieu]{vu2021structural}
Vu,~T.~V.; Kartamyshev,~A.; Hieu,~N.~V.; Dang,~T.~D.; Nguyen,~S.-N.;
  Poklonski,~N.; Nguyen,~C.~V.; Phuc,~H.~V.; Hieu,~N.~N. Structural, elastic,
  and electronic properties of chemically functionalized boron phosphide
  monolayer. \emph{RSC Advances} \textbf{2021}, \emph{11}, 8552--8558\relax
\mciteBstWouldAddEndPuncttrue
\mciteSetBstMidEndSepPunct{\mcitedefaultmidpunct}
{\mcitedefaultendpunct}{\mcitedefaultseppunct}\relax
\EndOfBibitem
\bibitem[Mohanta \latin{et~al.}(2019)Mohanta, Rawat, Jena, Dimple, Ahammed, and
  De~Sarkar]{mohanta2019interfacing}
Mohanta,~M.~K.; Rawat,~A.; Jena,~N.; Dimple,; Ahammed,~R.; De~Sarkar,~A.
  Interfacing boron monophosphide with molybdenum disulfide for an ultrahigh
  performance in thermoelectrics, two-dimensional excitonic solar cells, and
  nanopiezotronics. \emph{ACS Applied Materials \& Interfaces} \textbf{2019},
  \emph{12}, 3114--3126\relax
\mciteBstWouldAddEndPuncttrue
\mciteSetBstMidEndSepPunct{\mcitedefaultmidpunct}
{\mcitedefaultendpunct}{\mcitedefaultseppunct}\relax
\EndOfBibitem
\bibitem[Padavala \latin{et~al.}(2016)Padavala, Frye, Wang, Ding, Chen, Dudley,
  Raghothamachar, Lu, Flanders, and Edgar]{padavala2016epitaxy}
Padavala,~B.; Frye,~C.; Wang,~X.; Ding,~Z.; Chen,~R.; Dudley,~M.;
  Raghothamachar,~B.; Lu,~P.; Flanders,~B.; Edgar,~J. Epitaxy of boron
  phosphide on aluminum nitride (0001)/sapphire substrate. \emph{Crystal Growth
  \& Design} \textbf{2016}, \emph{16}, 981--987\relax
\mciteBstWouldAddEndPuncttrue
\mciteSetBstMidEndSepPunct{\mcitedefaultmidpunct}
{\mcitedefaultendpunct}{\mcitedefaultseppunct}\relax
\EndOfBibitem
\bibitem[Wiensch \latin{et~al.}(2017)Wiensch, John, Velazquez, Torelli,
  Pieterick, McDowell, Sun, Zhao, Brunschwig, and
  Lewis]{wiensch2017comparative}
Wiensch,~J.~D.; John,~J.; Velazquez,~J.~M.; Torelli,~D.~A.; Pieterick,~A.~P.;
  McDowell,~M.~T.; Sun,~K.; Zhao,~X.; Brunschwig,~B.~S.; Lewis,~N.~S.
  Comparative study in acidic and alkaline media of the effects of pH and
  crystallinity on the hydrogen-evolution reaction on MoS2 and MoSe2. \emph{ACS
  Energy Letters} \textbf{2017}, \emph{2}, 2234--2238\relax
\mciteBstWouldAddEndPuncttrue
\mciteSetBstMidEndSepPunct{\mcitedefaultmidpunct}
{\mcitedefaultendpunct}{\mcitedefaultseppunct}\relax
\EndOfBibitem
\bibitem[Martin(2004)]{martin2004electronic}
Martin,~R.~M. \emph{Electronic structure: basic theory and practical methods};
  Cambridge University Press, 2004\relax
\mciteBstWouldAddEndPuncttrue
\mciteSetBstMidEndSepPunct{\mcitedefaultmidpunct}
{\mcitedefaultendpunct}{\mcitedefaultseppunct}\relax
\EndOfBibitem
\bibitem[Martin \latin{et~al.}(2016)Martin, Reining, and
  Ceperley]{martin2016interacting}
Martin,~R.~M.; Reining,~L.; Ceperley,~D.~M. \emph{Interacting Electrons};
  Cambridge University Press, 2016\relax
\mciteBstWouldAddEndPuncttrue
\mciteSetBstMidEndSepPunct{\mcitedefaultmidpunct}
{\mcitedefaultendpunct}{\mcitedefaultseppunct}\relax
\EndOfBibitem
\bibitem[Freysoldt \latin{et~al.}(2014)Freysoldt, Grabowski, Hickel,
  Neugebauer, Kresse, Janotti, and Van~de Walle]{freysoldt2014RevModPhys}
Freysoldt,~C.; Grabowski,~B.; Hickel,~T.; Neugebauer,~J.; Kresse,~G.;
  Janotti,~A.; Van~de Walle,~C.~G. First-principles calculations for point
  defects in solids. \emph{Reviews of Modern Physics} \textbf{2014}, \emph{86},
  253\relax
\mciteBstWouldAddEndPuncttrue
\mciteSetBstMidEndSepPunct{\mcitedefaultmidpunct}
{\mcitedefaultendpunct}{\mcitedefaultseppunct}\relax
\EndOfBibitem
\bibitem[Feng \latin{et~al.}(2014)Feng, Su, Chen, and
  Liu]{feng2014MaterChemPhys}
Feng,~L.-p.; Su,~J.; Chen,~S.; Liu,~Z.-t. First-principles investigations on
  vacancy formation and electronic structures of monolayer MoS{$_2$}.
  \emph{Materials Chemistry and Physics} \textbf{2014}, \emph{148}, 5--9\relax
\mciteBstWouldAddEndPuncttrue
\mciteSetBstMidEndSepPunct{\mcitedefaultmidpunct}
{\mcitedefaultendpunct}{\mcitedefaultseppunct}\relax
\EndOfBibitem
\bibitem[Hohenberg and Kohn(1964)Hohenberg, and Kohn]{hohenberg1964PhysRev}
Hohenberg,~P.; Kohn,~W. Inhomogeneous electron gas. \emph{Physical Review}
  \textbf{1964}, \emph{136}, B864\relax
\mciteBstWouldAddEndPuncttrue
\mciteSetBstMidEndSepPunct{\mcitedefaultmidpunct}
{\mcitedefaultendpunct}{\mcitedefaultseppunct}\relax
\EndOfBibitem
\bibitem[Kohn and Sham(1965)Kohn, and Sham]{kohn1965PhysRev}
Kohn,~W.; Sham,~L.~J. Self-consistent equations including exchange and
  correlation effects. \emph{Physical Review} \textbf{1965}, \emph{140},
  A1133\relax
\mciteBstWouldAddEndPuncttrue
\mciteSetBstMidEndSepPunct{\mcitedefaultmidpunct}
{\mcitedefaultendpunct}{\mcitedefaultseppunct}\relax
\EndOfBibitem
\bibitem[Kresse and Furthm{\"u}ller(1996)Kresse, and
  Furthm{\"u}ller]{kresse1996efficient}
Kresse,~G.; Furthm{\"u}ller,~J. Efficient iterative schemes for ab initio
  total-energy calculations using a plane-wave basis set. \emph{Physical Review
  B} \textbf{1996}, \emph{54}, 11169\relax
\mciteBstWouldAddEndPuncttrue
\mciteSetBstMidEndSepPunct{\mcitedefaultmidpunct}
{\mcitedefaultendpunct}{\mcitedefaultseppunct}\relax
\EndOfBibitem
\bibitem[Bl{\"o}chl(1994)]{blochl1994projector}
Bl{\"o}chl,~P.~E. Projector augmented-wave method. \emph{Physical Review B}
  \textbf{1994}, \emph{50}, 17953\relax
\mciteBstWouldAddEndPuncttrue
\mciteSetBstMidEndSepPunct{\mcitedefaultmidpunct}
{\mcitedefaultendpunct}{\mcitedefaultseppunct}\relax
\EndOfBibitem
\bibitem[Blum \latin{et~al.}(2009)Blum, Gehrke, Hanke, Havu, Havu, Ren, Reuter,
  and Scheffler]{blum2009ComputPhysCommun}
Blum,~V.; Gehrke,~R.; Hanke,~F.; Havu,~P.; Havu,~V.; Ren,~X.; Reuter,~K.;
  Scheffler,~M. Ab initio molecular simulations with numeric atom-centered
  orbitals. \emph{Computer Physics Communications} \textbf{2009}, \emph{180},
  2175--2196\relax
\mciteBstWouldAddEndPuncttrue
\mciteSetBstMidEndSepPunct{\mcitedefaultmidpunct}
{\mcitedefaultendpunct}{\mcitedefaultseppunct}\relax
\EndOfBibitem
\bibitem[Stampfl and Van~de Walle(1999)Stampfl, and Van~de
  Walle]{stampfl1999PRB}
Stampfl,~C.; Van~de Walle,~C. Density-functional calculations for III-V
  nitrides using the local-density approximation and the generalized gradient
  approximation. \emph{Physical Review B} \textbf{1999}, \emph{59}, 5521\relax
\mciteBstWouldAddEndPuncttrue
\mciteSetBstMidEndSepPunct{\mcitedefaultmidpunct}
{\mcitedefaultendpunct}{\mcitedefaultseppunct}\relax
\EndOfBibitem
\bibitem[Perdew \latin{et~al.}(1996)Perdew, Burke, and
  Ernzerhof]{perdew1996PRL}
Perdew,~J.~P.; Burke,~K.; Ernzerhof,~M. Generalized gradient approximation made
  simple. \emph{Physical Review Letters} \textbf{1996}, \emph{77}, 3865\relax
\mciteBstWouldAddEndPuncttrue
\mciteSetBstMidEndSepPunct{\mcitedefaultmidpunct}
{\mcitedefaultendpunct}{\mcitedefaultseppunct}\relax
\EndOfBibitem
\bibitem[Tkatchenko and Scheffler(2009)Tkatchenko, and
  Scheffler]{tkatchenko2009PRL}
Tkatchenko,~A.; Scheffler,~M. Accurate molecular van der Waals interactions
  from ground-state electron density and free-atom reference data.
  \emph{Physical Review Letters} \textbf{2009}, \emph{102}, 073005\relax
\mciteBstWouldAddEndPuncttrue
\mciteSetBstMidEndSepPunct{\mcitedefaultmidpunct}
{\mcitedefaultendpunct}{\mcitedefaultseppunct}\relax
\EndOfBibitem
\bibitem[Tkatchenko \latin{et~al.}(2012)Tkatchenko, DiStasio~Jr, Car, and
  Scheffler]{tkatchenko2012PRL}
Tkatchenko,~A.; DiStasio~Jr,~R.~A.; Car,~R.; Scheffler,~M. Accurate and
  efficient method for many-body van der Waals interactions. \emph{Physical
  Review Letters} \textbf{2012}, \emph{108}, 236402\relax
\mciteBstWouldAddEndPuncttrue
\mciteSetBstMidEndSepPunct{\mcitedefaultmidpunct}
{\mcitedefaultendpunct}{\mcitedefaultseppunct}\relax
\EndOfBibitem
\bibitem[Henkelman \latin{et~al.}(2000)Henkelman, Uberuaga, and
  J{\'o}nsson]{henkelman2000climbing}
Henkelman,~G.; Uberuaga,~B.~P.; J{\'o}nsson,~H. A climbing image nudged elastic
  band method for finding saddle points and minimum energy paths. \emph{The
  Journal of Chemical Physics} \textbf{2000}, \emph{113}, 9901--9904\relax
\mciteBstWouldAddEndPuncttrue
\mciteSetBstMidEndSepPunct{\mcitedefaultmidpunct}
{\mcitedefaultendpunct}{\mcitedefaultseppunct}\relax
\EndOfBibitem
\bibitem[Henkelman and J{\'o}nsson(2000)Henkelman, and
  J{\'o}nsson]{henkelman2000improved}
Henkelman,~G.; J{\'o}nsson,~H. Improved tangent estimate in the nudged elastic
  band method for finding minimum energy paths and saddle points. \emph{The
  Journal of Chemical Physics} \textbf{2000}, \emph{113}, 9978--9985\relax
\mciteBstWouldAddEndPuncttrue
\mciteSetBstMidEndSepPunct{\mcitedefaultmidpunct}
{\mcitedefaultendpunct}{\mcitedefaultseppunct}\relax
\EndOfBibitem
\bibitem[Fu \latin{et~al.}(2016)Fu, Luo, Li, and Yang]{Direct_Z_Fu}
Fu,~C.-F.; Luo,~Q.; Li,~X.; Yang,~J. Two-dimensional van der Waals
  nanocomposites as Z-scheme type photocatalysts for hydrogen production from
  overall water splitting. \emph{Journal of Materials Chemistry A}
  \textbf{2016}, \emph{4}, 18892--18898\relax
\mciteBstWouldAddEndPuncttrue
\mciteSetBstMidEndSepPunct{\mcitedefaultmidpunct}
{\mcitedefaultendpunct}{\mcitedefaultseppunct}\relax
\EndOfBibitem
\bibitem[Weng and Gao(2018)Weng, and Gao]{weng2018honeycomb}
Weng,~J.; Gao,~S.-P. A honeycomb-like monolayer of HfO 2 and the calculation of
  static dielectric constant eliminating the effect of vacuum spacing.
  \emph{Physical Chemistry Chemical Physics} \textbf{2018}, \emph{20},
  26453--26462\relax
\mciteBstWouldAddEndPuncttrue
\mciteSetBstMidEndSepPunct{\mcitedefaultmidpunct}
{\mcitedefaultendpunct}{\mcitedefaultseppunct}\relax
\EndOfBibitem
\bibitem[Ren \latin{et~al.}(2020)Ren, Tang, Sun, Cai, Cheng, and
  Zhang]{ren2020direct}
Ren,~K.; Tang,~W.; Sun,~M.; Cai,~Y.; Cheng,~Y.; Zhang,~G. A direct Z-scheme PtS
  2/arsenene van der Waals heterostructure with high photocatalytic water
  splitting efficiency. \emph{Nanoscale} \textbf{2020}, \emph{12},
  17281--17289\relax
\mciteBstWouldAddEndPuncttrue
\mciteSetBstMidEndSepPunct{\mcitedefaultmidpunct}
{\mcitedefaultendpunct}{\mcitedefaultseppunct}\relax
\EndOfBibitem
\bibitem[Rahman \latin{et~al.}(2018)Rahman, Morbec, Rahman, and
  Kratzer]{rahman2018commensurate}
Rahman,~A.~U.; Morbec,~J.~M.; Rahman,~G.; Kratzer,~P. Commensurate versus
  incommensurate heterostructures of group-III monochalcogenides.
  \emph{Physical Review Materials} \textbf{2018}, \emph{2}, 094002\relax
\mciteBstWouldAddEndPuncttrue
\mciteSetBstMidEndSepPunct{\mcitedefaultmidpunct}
{\mcitedefaultendpunct}{\mcitedefaultseppunct}\relax
\EndOfBibitem
\bibitem[Ren \latin{et~al.}(2019)Ren, Sun, Luo, Wang, Yu, and
  Tang]{ren2019first}
Ren,~K.; Sun,~M.; Luo,~Y.; Wang,~S.; Yu,~J.; Tang,~W. First-principle study of
  electronic and optical properties of two-dimensional materials-based
  heterostructures based on transition metal dichalcogenides and boron
  phosphide. \emph{Applied Surface Science} \textbf{2019}, \emph{476},
  70--75\relax
\mciteBstWouldAddEndPuncttrue
\mciteSetBstMidEndSepPunct{\mcitedefaultmidpunct}
{\mcitedefaultendpunct}{\mcitedefaultseppunct}\relax
\EndOfBibitem
\bibitem[He \latin{et~al.}(2018)He, Chen, Chen, Liu, Zhou, Li, and
  Wang]{he2018electrochemical}
He,~Q.; Chen,~X.; Chen,~S.; Liu,~L.; Zhou,~F.; Li,~X.-B.; Wang,~G.
  Electrochemical hydrogen evolution at the interface of monolayer VS2 and
  water from first-principles calculations. \emph{ACS Applied Materials \&
  Interfaces} \textbf{2018}, \emph{11}, 2944--2949\relax
\mciteBstWouldAddEndPuncttrue
\mciteSetBstMidEndSepPunct{\mcitedefaultmidpunct}
{\mcitedefaultendpunct}{\mcitedefaultseppunct}\relax
\EndOfBibitem
\end{mcitethebibliography}


\providecommand{\latin}[1]{#1}
\makeatletter
\providecommand{\doi}
  {\begingroup\let\do\@makeother\dospecials
  \catcode`\{=1 \catcode`\}=2 \doi@aux}
\providecommand{\doi@aux}[1]{\endgroup\texttt{#1}}
\makeatother
\providecommand*\mcitethebibliography{\thebibliography}
\csname @ifundefined\endcsname{endmcitethebibliography}
  {\let\endmcitethebibliography\endthebibliography}{}
\begin{mcitethebibliography}{3}
\providecommand*\natexlab[1]{#1}
\providecommand*\mciteSetBstSublistMode[1]{}
\providecommand*\mciteSetBstMaxWidthForm[2]{}
\providecommand*\mciteBstWouldAddEndPuncttrue
  {\def\EndOfBibitem{\unskip.}}
\providecommand*\mciteBstWouldAddEndPunctfalse
  {\let\EndOfBibitem\relax}
\providecommand*\mciteSetBstMidEndSepPunct[3]{}
\providecommand*\mciteSetBstSublistLabelBeginEnd[3]{}
\providecommand*\EndOfBibitem{}
\mciteSetBstSublistMode{f}
\mciteSetBstMaxWidthForm{subitem}{(\alph{mcitesubitemcount})}
\mciteSetBstSublistLabelBeginEnd
  {\mcitemaxwidthsubitemform\space}
  {\relax}
  {\relax}

\bibitem[Onida \latin{et~al.}(2002)Onida, Reining, and
  Rubio]{onida2002electronic}
Onida,~G.; Reining,~L.; Rubio,~A. Electronic excitations: density-functional
  versus many-body Green’s-function approaches. \emph{Reviews of Modern
  Physics} \textbf{2002}, \emph{74}, 601\relax
\mciteBstWouldAddEndPuncttrue
\mciteSetBstMidEndSepPunct{\mcitedefaultmidpunct}
{\mcitedefaultendpunct}{\mcitedefaultseppunct}\relax
\EndOfBibitem
\bibitem[Jiang \latin{et~al.}(2012)Jiang, Rinke, and
  Scheffler]{jiang2012electronic}
Jiang,~H.; Rinke,~P.; Scheffler,~M. Electronic properties of lanthanide oxides
  from the G W perspective. \emph{Physical Review B} \textbf{2012}, \emph{86},
  125115\relax
\mciteBstWouldAddEndPuncttrue
\mciteSetBstMidEndSepPunct{\mcitedefaultmidpunct}
{\mcitedefaultendpunct}{\mcitedefaultseppunct}\relax
\EndOfBibitem
\end{mcitethebibliography}
\newpage


\end{document}









\begin{center}
	{\Large \bf Supporting Information}\\ 
\end{center}
\vspace*{12pt}

\noindent {\bf \Large I. Structural, Electronic and Optical Properties of MoS{$_2$}/BP and MoSSe/BP\\}
\noindent {\bf \Large II. Planar Averaged Charge Density Plot of MoS{$_2$}/BP and MoSSe/BP\\}
\noindent {\bf \Large III. Planar Averaged Charge Density Plot of MoS{$_2$}/BP with 1 Solvated H{$^+$}\\}
\noindent {\bf \Large IV. Radial Distribution Function at 0K and 300K Temperatures for MoS{$_2$}/BP Configuration\\}
\noindent {\bf \Large V. $3\times3$ MoS{$_2$}/BP Configurations\\}
\noindent {\bf \Large VI. $3\times3$ MoS{$_2$}/BP Configurations with H$_\textrm{down}$ Water Layer Orientation\\}
\noindent {\bf \Large VII. Electrode Potential of $2\times2$ MoS{$_2$}/BP and MoSSe/BP\\}
\noindent {\bf \Large VIII. H$_\textrm{down}$ Water Orientation Reaction Path and Electrode Potential\\}
\noindent {\bf \Large IX. Tafel Reaction Steps for $2\times2$ MoS{$_2$}/BP in case of 4H{$_2$}O}

\newpage
\section{\bf \Large Structural, Electronic and Optical Properties of MoS{$_2$}/BP and MoSSe/BP}
\vspace*{2cm}
\begin{figure}[!h]
	\centering
	\includegraphics[width=0.75\columnwidth,clip]{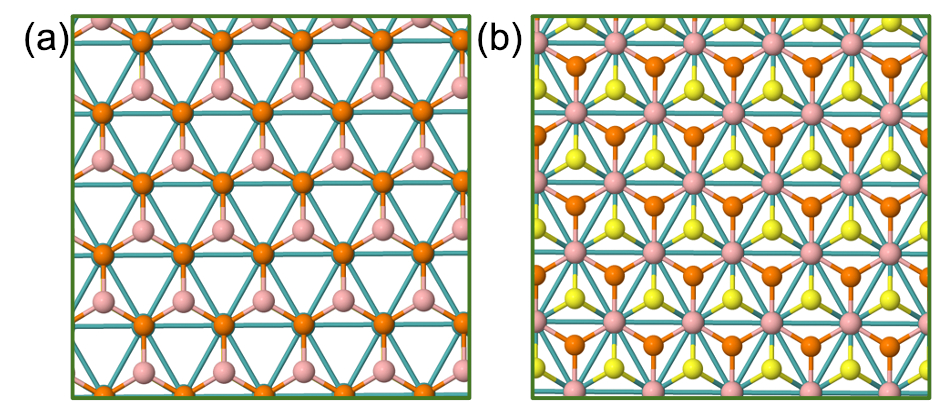}
	\caption{(Color online) (a) and (b) are the two stacking styles of MoS{$_2$}/BP analyzed in our study. Configuration (b) has minimum binding energy.}
	\label{fig:Stack}
\end{figure}
\begin{figure}[!h]
	\centering
	\includegraphics[width=0.75\columnwidth,clip]{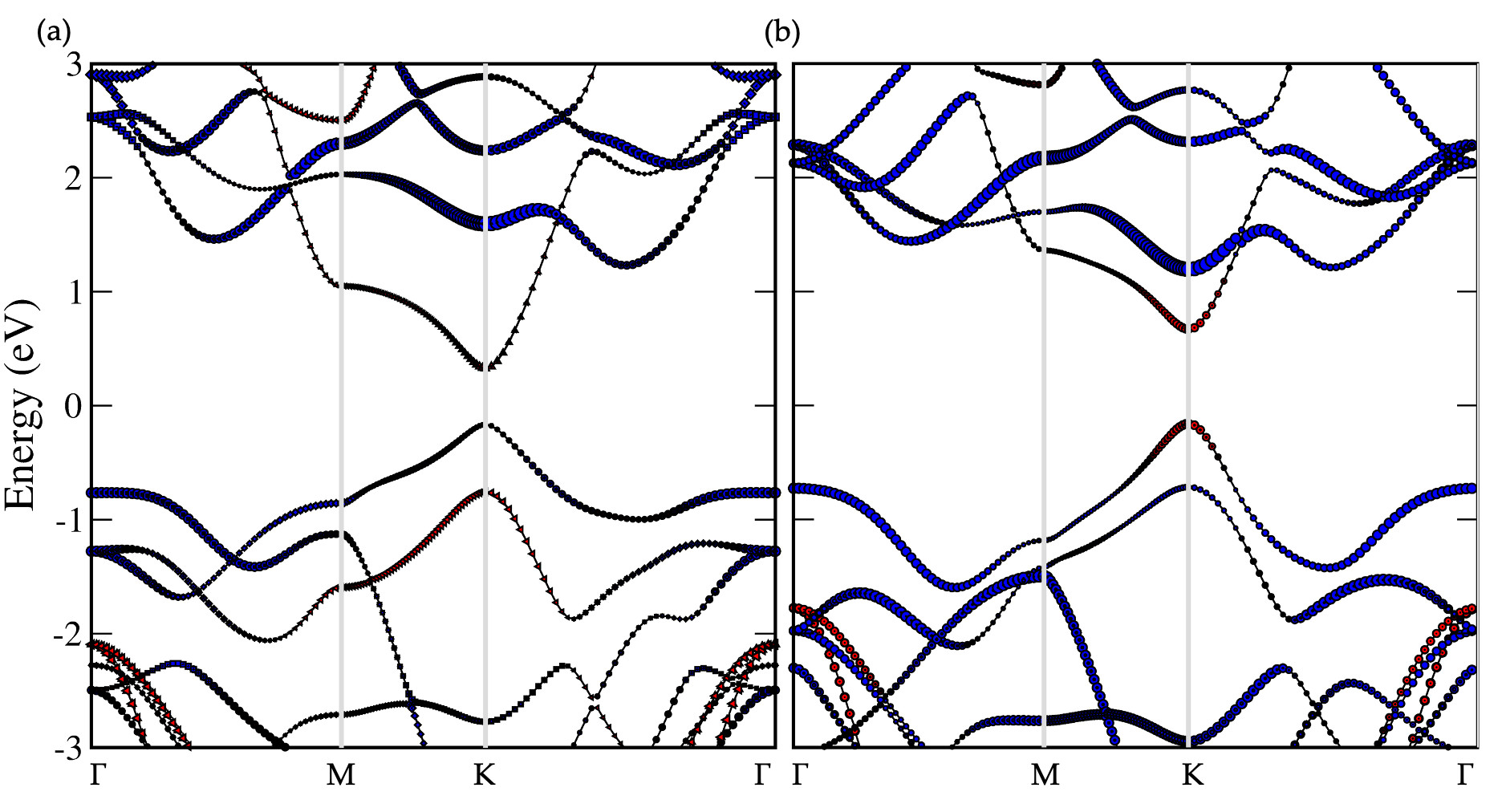}
	\caption{(Color online) (a) Type II bandstructure obtained for MoSSe/BP vdW HTS; (b) Type I bandstructure obtained for MoS{$_2$}/BP vdW HTS. (xc functional: PBE)}
	\label{fig:Bandstructure}
\end{figure}
\noindent In case of MoSSe/BP, BP contributes to the CBm (Conduction Band Minimum) and MoSSe contributes to the VBM (Valence Band Maximum). MoS{$_2$}/BP shows type 1 band edge alignment where MoS{$_2$} straddles BP band edges.
\newpage
\begin{figure}[h]
	\centering
	\includegraphics[width=0.75\columnwidth,clip]{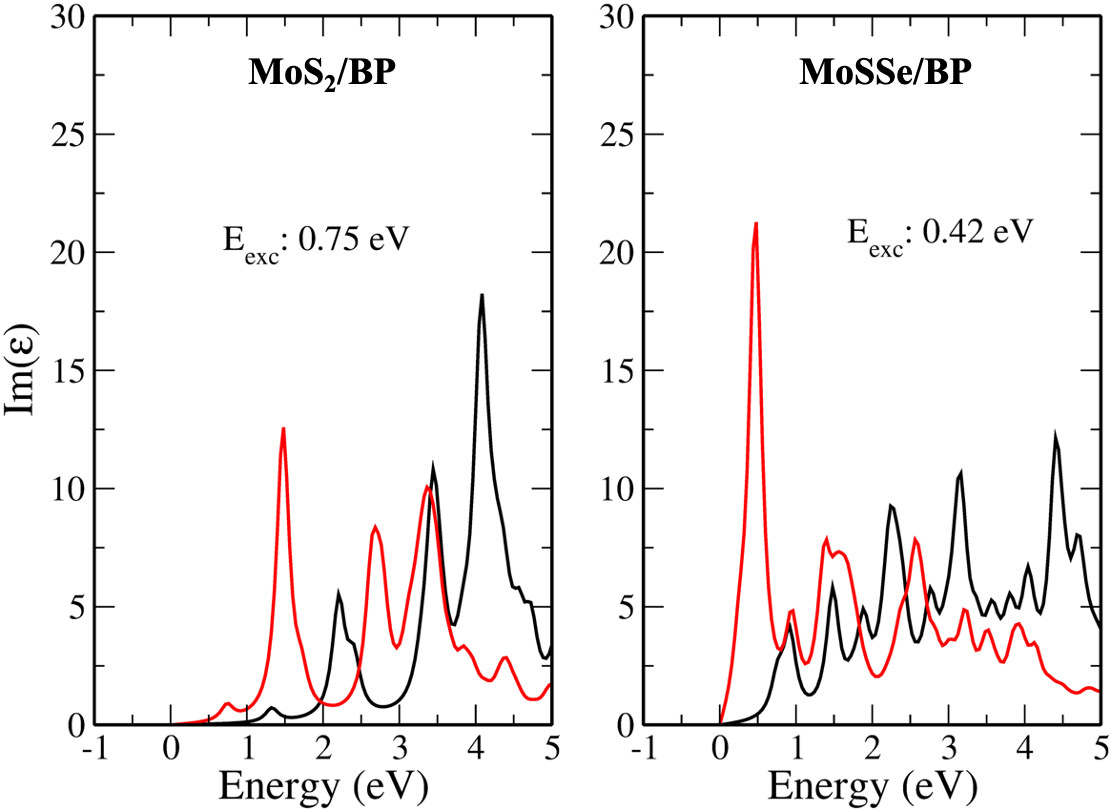}
	\caption{(Color online) Exciton binding energy obtained for MoS{$_2$}/BP and MoSSe/BP vdW HTS.}
	\label{fig:5}
\end{figure}
\noindent The optical properties have also been calculated by the GW approach (Many Body Perturbation Theory (MBPT)). Fig. S3 shows the optical response using GW@BSE method that calculates the dielectric function. This is a complex function where the expression for interband process is the imaginary part thereby giving the absorption spectra. The real part (Re($\epsilon$)) is deduced from the Kramers-Kronig relation. We observe both the vdW HTSs (MoS{$_2$}/BP and MoSSe/BP) having response in the visible region. We have calculated the exciton binding energy (E{$_\textrm{B}$}) using the Bethe-Salpeter Equation. We observe small E{$_\textrm{B}$} that indicates its applicability in photoelectrochemical processes.\\
\newpage
\section{\bf \Large Planar Averaged Charge Density Plot of MoS{$_2$}/BP and MoSSe/BP}
\begin{figure}[!h]
	\centering
	\includegraphics[width=0.75\columnwidth,clip]{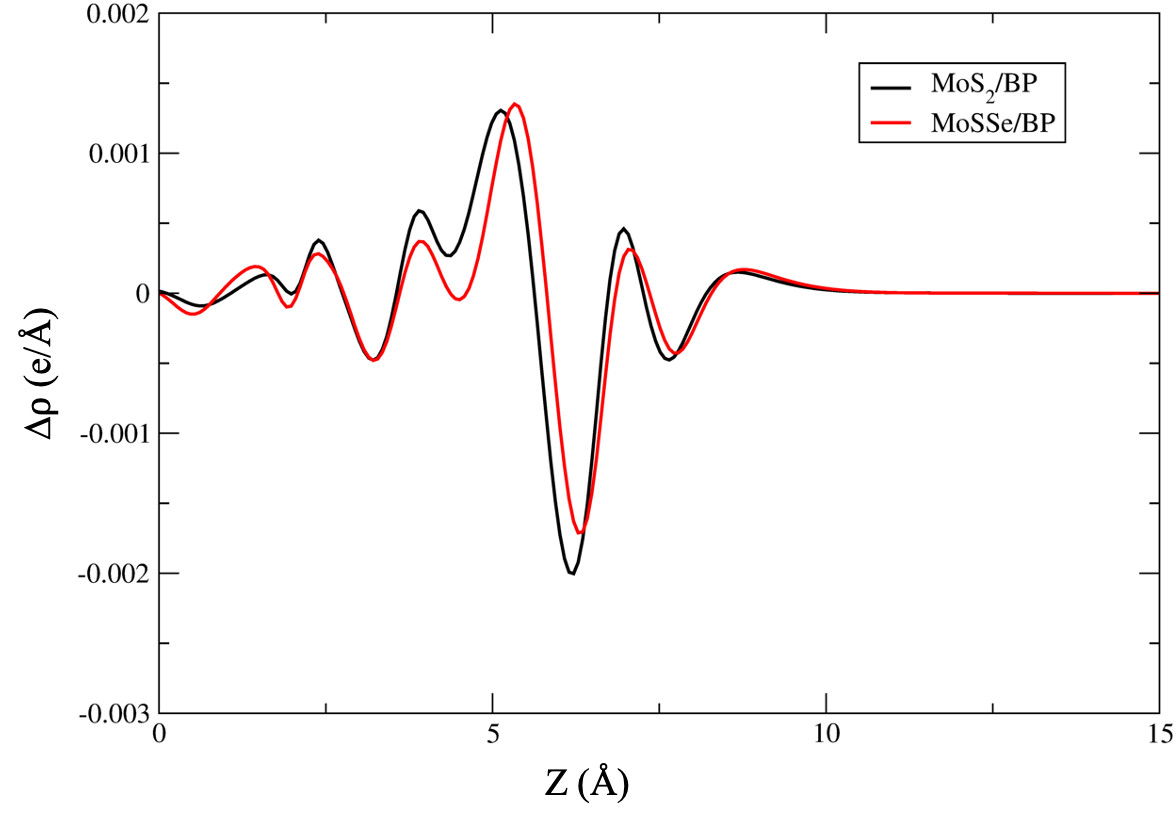}
	\caption{(Color online) Planar averaged charge density plot for MoS{$_2$}/BP and MoSSe/BP vdW HTS.}
	\label{fig:Charge}
\end{figure}

\noindent The planar averaged charge density difference $\Delta \rho$ is calculated by:
\begin{equation}
	\Delta \rho = \rho(\textrm{vdW HTSs}) - \rho(\textrm{MoS}_2 \textrm{ or MoSSe}) - \rho(\textrm{BP})
\end{equation}
\noindent where {$ \rho(\textrm{vdW HTSs})$}, {$ \rho(\textrm{MoS}_2) $}, {$ \rho(\textrm{MoSSe}) $} and {$ \rho(\textrm{BP}) $}  are the charge densities of the vdW HTS, monolayer MoS{$_2$}, monolayer MoSSe and monolayer BP, respectively.
\newpage
\section{\bf \Large Planar Averaged Charge Density Plot of MoS{$_2$}/BP with 1 Solvated H{$^+$}}
\begin{figure}[!h]
	\centering
	\includegraphics[width=0.8\columnwidth,clip]{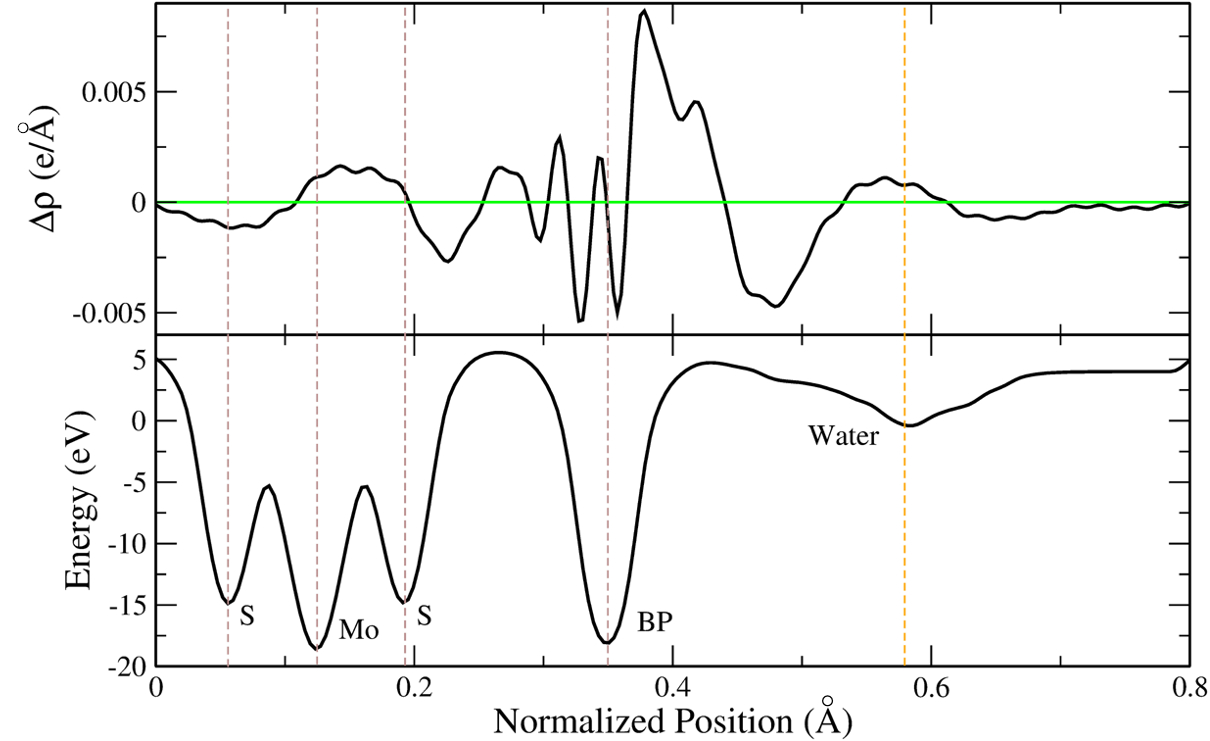}
	\caption{(Color online) Planar averaged charge density and electrostatic potential plot for MoS{$_2$}/BP (with 1 H{$^+$}).}
	\label{fig:4}
\end{figure}
\noindent Fig.~\ref{fig:4} shows the planar averaged charge density plot and the corresponding electrostatic potential plot for MoS{$_2$}/BP with 1 solvated H{$^+$}. The electrostatic potential plot explains the electrostatic potential corresponding to each atomic layer. We observe significant charge transfer at the BP and water interface.
\newpage
\section{\bf \Large Radial Distribution Function at 0K and 300K Temperatures for MoS{$_2$}/BP Configuration}
\begin{figure}[!h]
	\centering
	\includegraphics[width=0.8\columnwidth,clip]{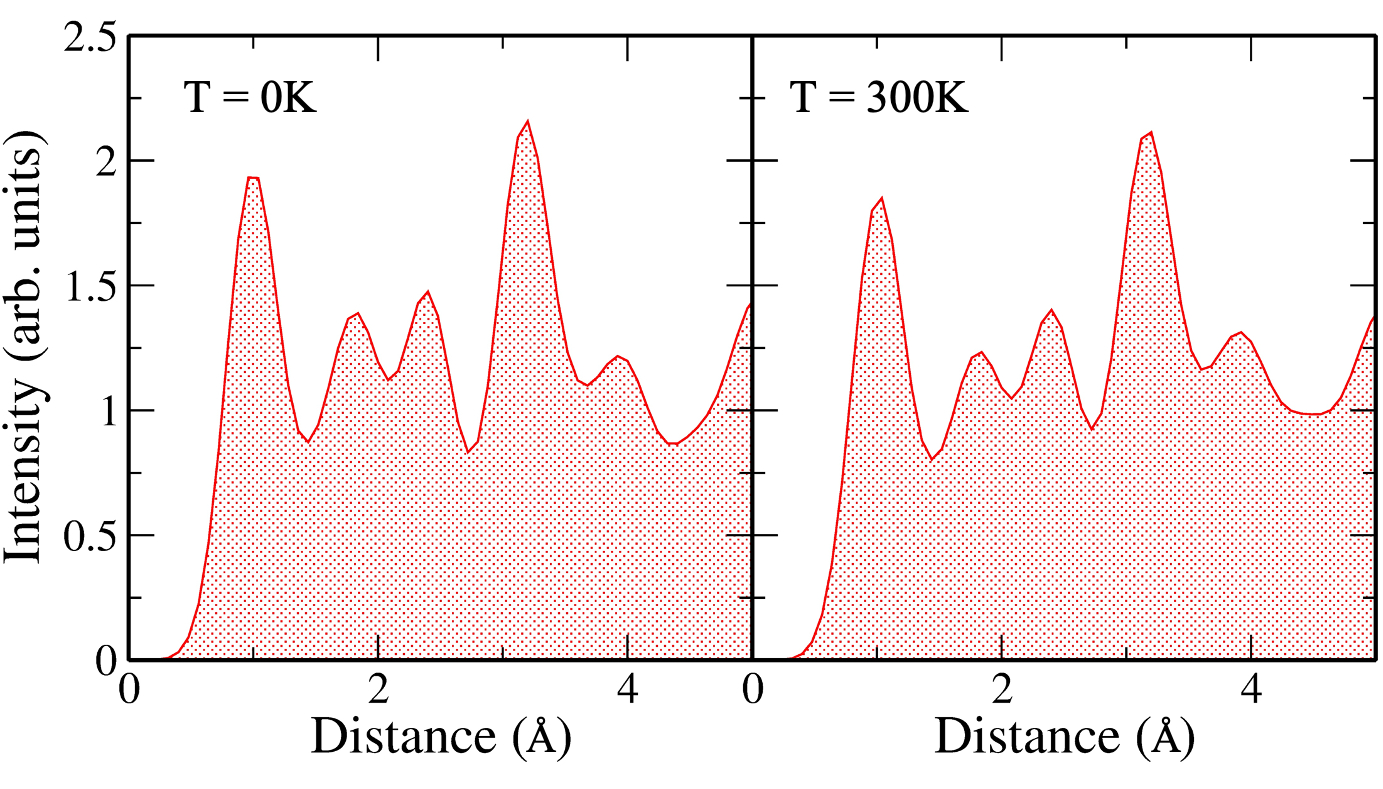}
	\caption{(Color online) Radial distribution function of MoS{$_2$}/BP at two different temperatures T= 0K and T= 300K.}
	\label{fig:rdf}
\end{figure}
\noindent The radial distribution function here indicates the structural similarity of the MoS{$_2$}/BP with water layer at 0K and 300K. The water orientation thus obtained corroborates with the initial optimized configuration.
\newpage
\section{\bf \Large $3\times3$ MoS{$_2$}/BP Configurations}
\begin{figure}[!h]
	\centering
	\includegraphics[width=0.6\columnwidth,clip]{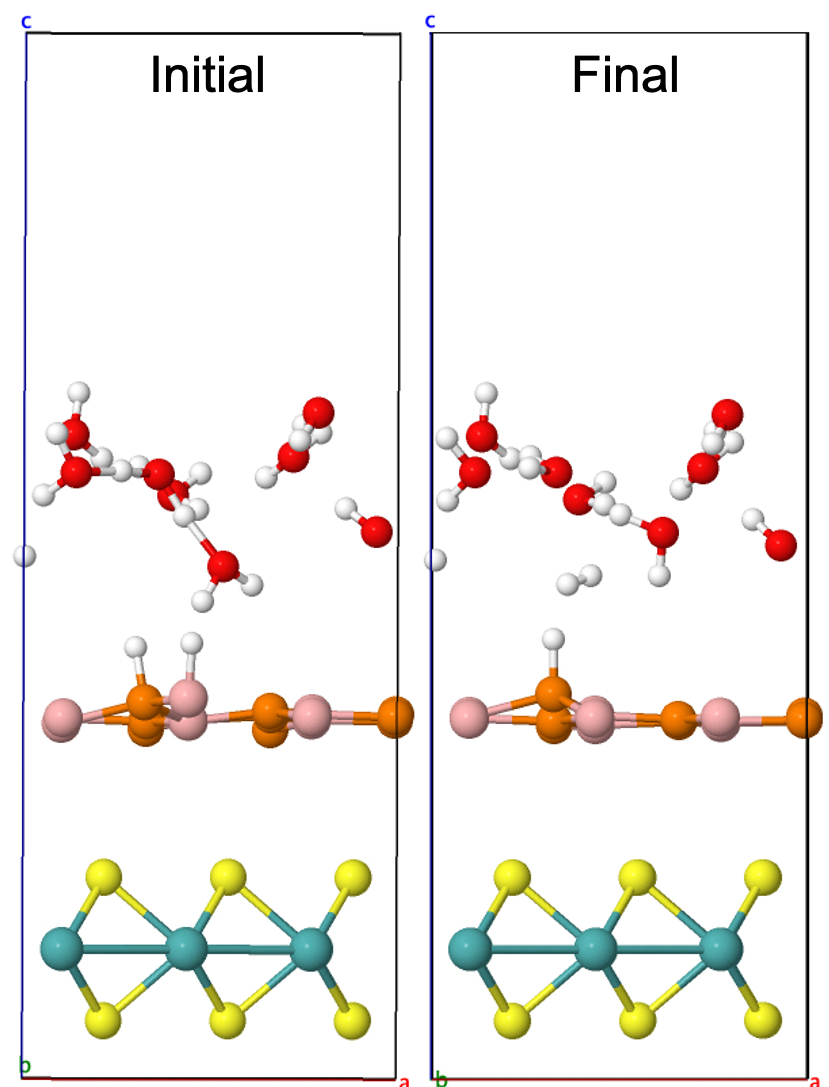}
	\caption{(Color online) Initial and final configuration of $3\times3$ MoS{$_2$}/BP for Heyrovsky reaction.}
	\label{fig:ini_fin}
\end{figure}
\newpage
\section{\bf \Large $3\times3$ MoS{$_2$}/BP Configurations With H$_\textrm{down}$ Water Layer Orientation}
\begin{figure}[!h]
	\centering
	\includegraphics[width=0.6\columnwidth,clip]{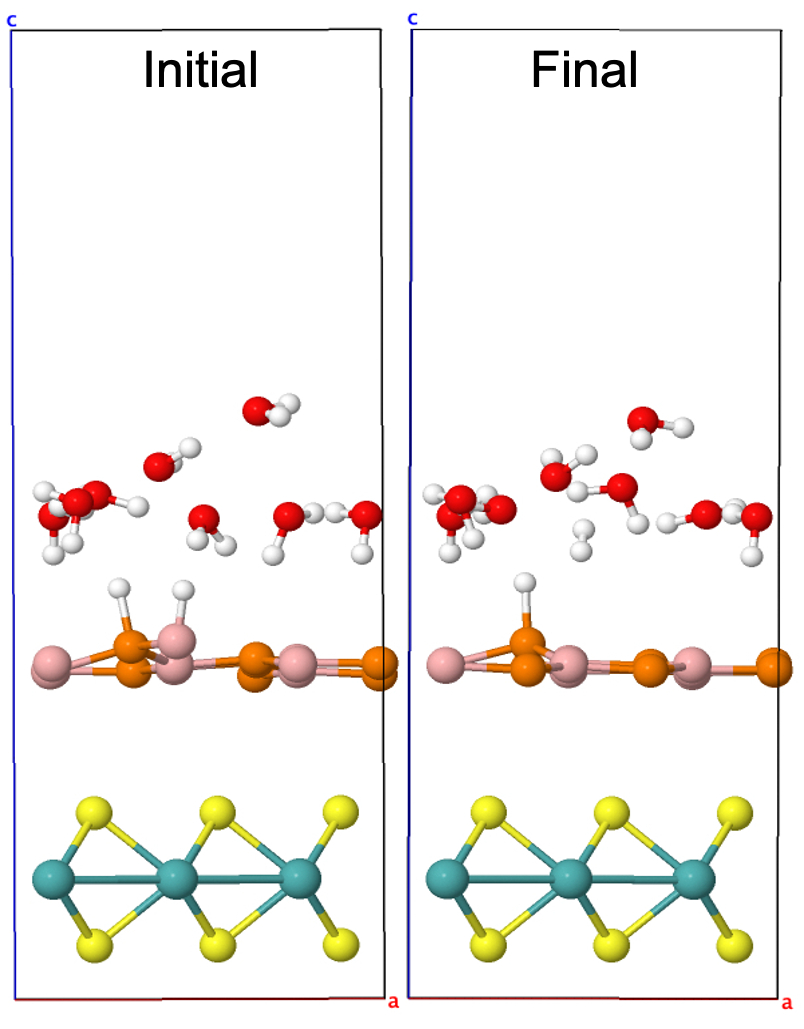}
	\caption{(Color online) Initial and final configuration of $3\times3$ MoS{$_2$}/BP for Heyrovsky reaction in case H$_\textrm{down}$ water layer orientation.}
	\label{fig:hdown_orient}
\end{figure}
\newpage
\section{\bf \Large Electrode Potential of $2\times2$ MoS{$_2$}/BP and MoSSe/BP}
\begin{figure}[!h]
	\centering
	\includegraphics[width=0.95\columnwidth,clip]{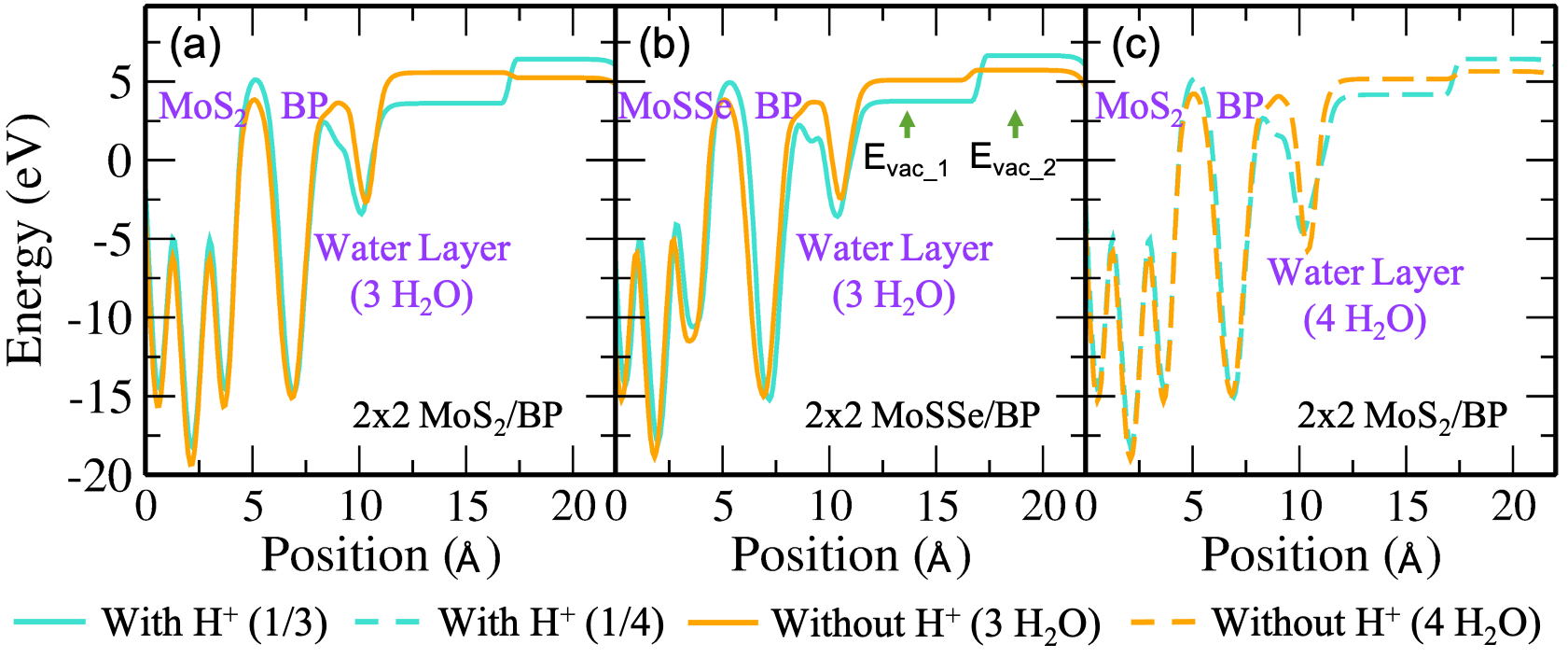}
	\caption{(Color online) (a) Electrode potential for $2\times2$ MoS{$_2$}/BP with and without H{$^+$} corresponding to 1/3 proton concentration and 3 H{$_2$}O, respectively.  (b) Electrode potential for $2\times2$ MoSSe/BP with and without H{$^+$} corresponding to 1/3 proton concentration and 3 H{$_2$}O, respectively. (c) Electrode potential for $2\times2$ MoS{$_2$}/BP with and without H{$^+$} corresponding to 1/4 proton concentration and 4 H{$_2$}O, respectively.}
	\label{fig:2x2}
\end{figure}
\newpage
\section{\bf \Large H$_\textrm{down}$ Water  Orientation Reaction Path and Electrode Potential}
\begin{figure}[!h]
	\centering
	\includegraphics[width=0.8\columnwidth,clip]{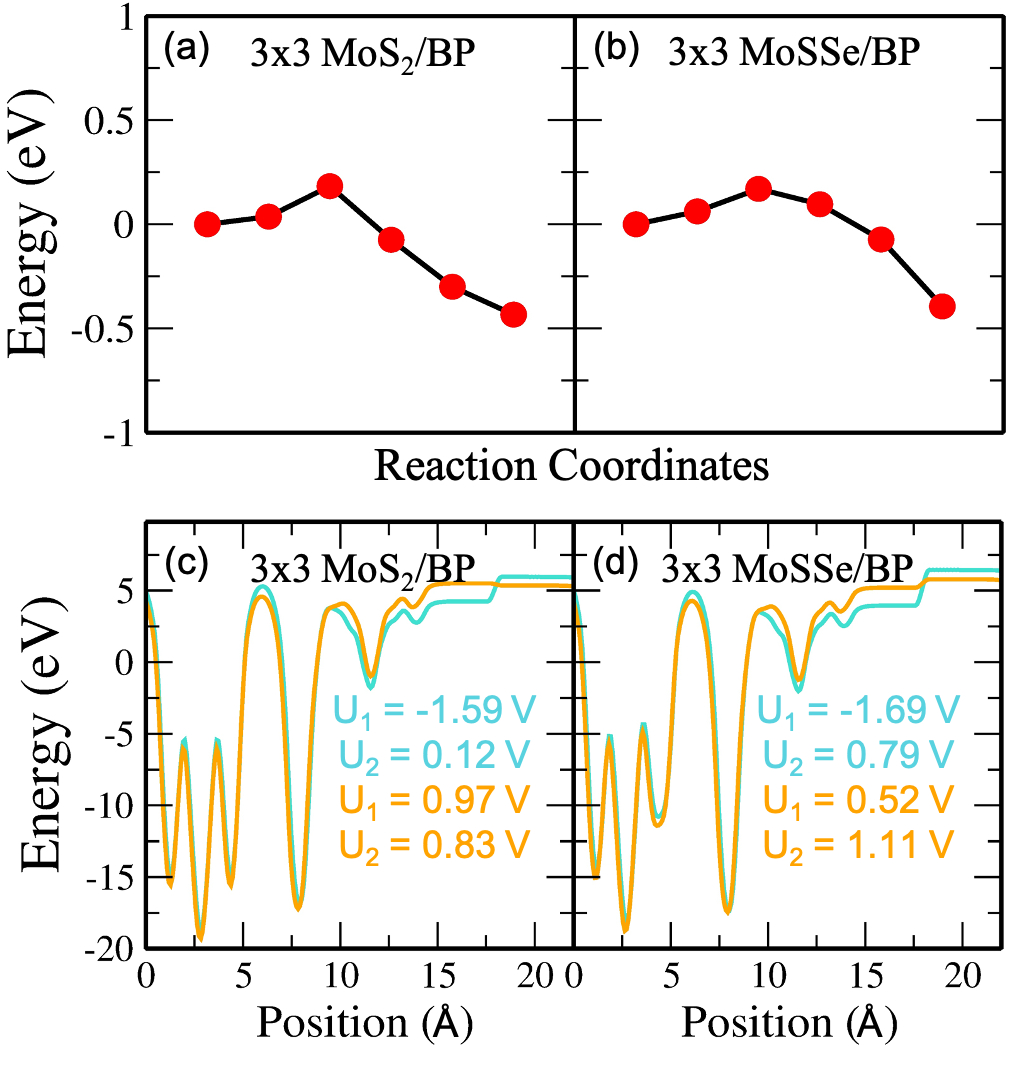}
	\caption{(Color online) (a) and (b) Heyrovsky reaction step for MoS{$_2$}/BP and MoSSe/BP vdW HTS in $3\times3$ supercell with 1/8 proton concentration (c) and (d) Electrostatic potential plot of MoS{$_2$}/BP and MoSSe/BP vdW HTSs depicting water layer with and without H{$^+$} in $3\times3$ supercell with 1/8 proton concentration and 8 water molecules, respectively.}
	\label{fig:hdown}
\end{figure}

\newpage
\section{\bf \Large  Tafel Reaction Steps for $2\times2$ MoS{$_2$}/BP in case of 4H{$_2$}O}
\begin{figure}[!h]
	\centering
	\includegraphics[width=0.55\columnwidth,clip]{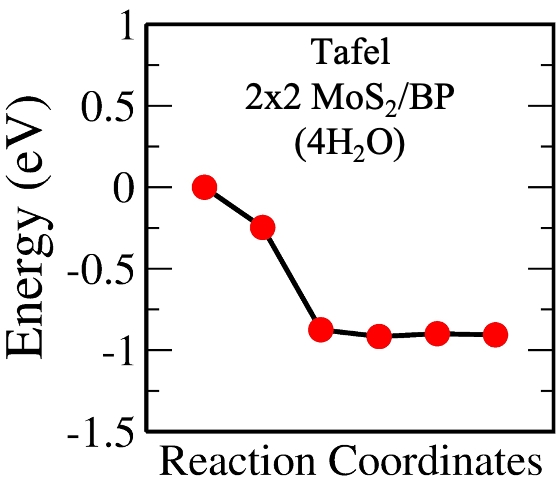}
	\caption{(Color online) Tafel reaction step for $2\times2$ MoS{$_2$}/BP for 4 H{$_2$}O.}
	\label{fig:2x2_1by4}
\end{figure}
\clearpage
\newpage